\documentclass[lettersize,journal]{IEEEtran}
\usepackage{amsmath,amsfonts}
\usepackage{array}
\usepackage[caption=false,font=footnotesize,labelfont=sf,textfont=sf]{subfig}
\usepackage{textcomp}
\usepackage{stfloats}
\usepackage{url}
\usepackage{verbatim}
\usepackage{graphicx}
\usepackage{cite}
\usepackage{balance}
\usepackage{hyperref}
\hypersetup{hidelinks}

\usepackage[linesnumbered,ruled,vlined]{algorithm2e}
\usepackage{mathtools}   
\usepackage{multirow}
\usepackage{enumitem}

\SetCommentSty{myCommentStyle}
\usepackage{booktabs} 
\usepackage{amsmath}
\usepackage{amssymb}

\begin{document}

\title{Oases: Efficient Large-Scale Model Training on Commodity Servers via Overlapped and Automated Tensor Model Parallelism}


\author{\IEEEauthorblockN{Shengwei Li, Zhiquan Lai, Dongsheng Li, Yanqi Hao, Weijie Liu, Keshi Ge, Xiaoge Deng, Kai Lu} \\
\thanks{Shengwei~Li, Zhiquan~Lai, Dongsheng~Li, Yanqi~Hao, Weijie~Liu, Keshi~Ge,  and Kai~Lu are with the National Key Laboratory of Parallel and Distributed Computing, College of Computer Science and Technology, National University of Defense Technology in Changsha, Hunan, China. Xiaoge~Deng is with the Intelligent Game and Decision Lab, Beijing, China. \\
E-mail: \{swli,zqlai,dsli,yqhao,liuweijie,gekeshi,kailu\}@nudt.edu.cn \\
Zhiquan Lai is the corresponding author of this paper.}
\thanks{This work was supported by the National Natural Science Foundation of China under Grant No. 62025208 and 62421002.}
}



\maketitle

\begin{abstract}
Deep learning is experiencing a rise in large-scale models. 
Training large-scale models is costly, prompting researchers to train large-scale models on commodity servers that more researchers can access. The massive number of parameters necessitates the use of model parallelism training methods. Existing studies focus on training with pipeline model parallelism. However, the tensor model parallelism (TMP) is inevitable when the model size keeps increasing, where frequent data-dependent communication and computation operations significantly reduce the training efficiency. 

In this paper, we present Oases, an automated TMP method with overlapped communication to accelerate large-scale model training on commodity servers. Oases proposes a fine-grained training operation schedule to maximize overlapping communication and computation that have data dependence. Additionally, we design the Oases planner that searches for the best model parameter partition strategy of TMP to achieve further accelerations. Unlike existing methods, Oases planner is tailored to model the cost of overlapped communication-computation operations. 

We evaluate Oases on various model settings and two commodity clusters, and compare Oases to four state-of-the-art implementations. Experimental results show that Oases achieves speedups of 1.01--1.48\(\times\) over the fastest baseline, and speedups of up to 1.95\(\times\) over Megatron.
\end{abstract}

\begin{IEEEkeywords}
Large-scale model, distributed training, tensor model parallelism, communication overlapping.
\end{IEEEkeywords}

\section{Introduction}
\IEEEPARstart{R}{ecent}
 works have demonstrated the success of transformer-based large-scale models~\cite{bommasani2021opportunities} in various downstream tasks. 
For example, large generative models~\cite{brown2020language} have enabled fundamentally new capabilities on copilot applications.
The accuracy of transformer-based models increases with its model size~\cite{kaplan2020scaling}, but training such models consumes a significant amount of time and resources~\cite{sharir2020cost}.

Unfortunately, only those who have access to massive datacenter-based resources are typically capable of undertaking the training of large-scale models.
Considering the cost-effectiveness of commodity GPUs (e.g., compared to A100, the RTX3090 GPU provides 58\% of computing performance at 1/10 price~\cite{vastprice}), they can train or fine-tune large models timely and not costly.
To democratize large-scale model training and make it accessible to a broader range of researchers, addressing the challenge of training large-scale models on cost-effective commodity servers emerges as a significant concern~\cite{li2022harmony,feng2023mobius,eliad2021fine}.

\begin{figure}

\centering
\includegraphics[width=\linewidth]{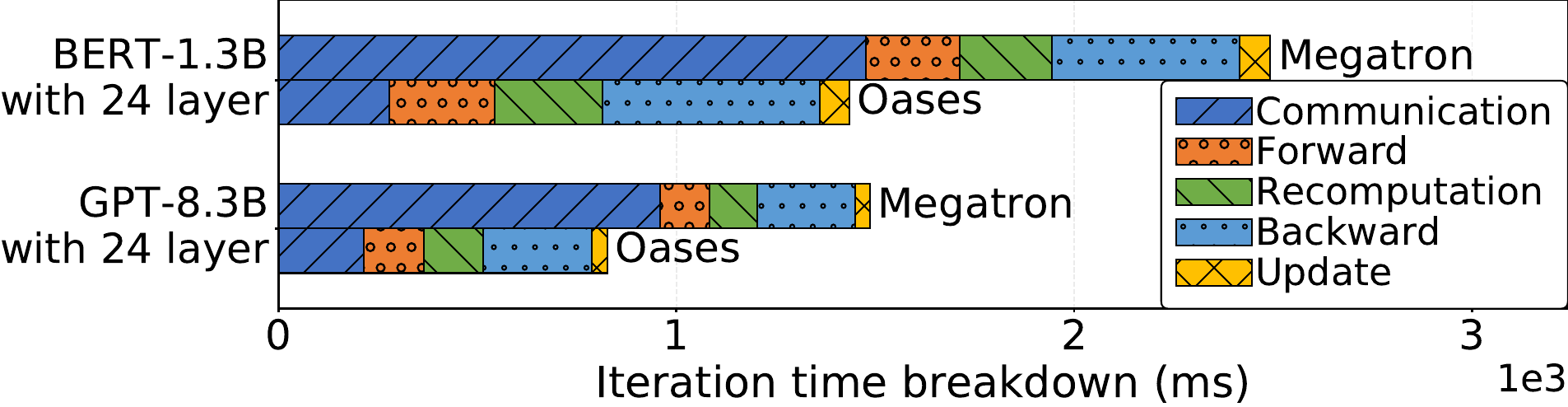}
\caption{TMP training iteration breakdown of two transformer-based models. }
\label{fig:motivation}
\end{figure}

Model-parallel training is necessary to accommodate large models in device memories, particularly on commodity hardware with constrained capacity. 
Existing training systems~\cite{li2022harmony,feng2023mobius,eliad2021fine} for training large-scale on commodity hardware are tailored to pipeline model parallelism (PMP), where model layers are scattered across devices. 
However, as model sizes increase, the memory cost of a single layer becomes significant, resulting in a deeper pipeline and decreased training efficiency. 
Designed for the transformer-based models, the tensor model parallelism (TMP) partitions the computation of specific operators into parallel devices~\cite{shoeybi2019megatron}, often used in combination with PMP for training large-scale models in datacenters.
TMP involves an extensive amount of blocked communication operations, which becomes the main performance bottleneck in large-scale model training on commodity servers, where communication bandwidth is limited. 
For instance, when training a large model on the commodity GPUs RTX3090 without NVLink, the communication overhead of TMP can reach 64.6\% (Fig.~\ref{fig:motivation}). 
Thus, it is indispensable to \textbf{design a TMP training system that improves both user accessibility and training efficiency on cost-effective servers}.

To alleviate the communication bottleneck of TMP,  the literature extensively explores two main lines of research: 1) overlapping operations, and 2) optimizing parallel strategies.
Communication-computation overlap through operation scheduling is a common technique to reduce the communication overhead in data parallelism~\cite{zhang2017poseidon,shi2019mg,peng2019generic,mahajanSYNDICATE}.
However, communication and computation are data-dependent in TMP training, making it challenging to exploit the full potential of overlapping. Existing approaches on TMP apply sub-optimal training schedules since they only overlap communication with computation within matrix multiplication operators\cite{wang2022overlap,zeng2022acctfm,jangda2022breaking} or single propagation pass~\cite{lai2023merak}.
Therefore, the first core problem this paper tries to answer is how to \textbf{optimize the TMP training operation schedule that maximizes overlapping communication and computation}.

Automated parallelization plays a crucial role in further harnessing the computational capabilities of commodity devices in large-scale model training.
However, finding an appropriate distributed training scheme for communication computation overlapping schedules is non-trivial, as it is difficult to estimate the cost of overlapped operations.
Previous studies can hardly give consideration to both overlapping and model parallel strategy searching on TMP, overlooking the potential of further training performance improvement.
Communication-computation overlap works lack automated partition strategy planning, while auto-parallelism approaches~\cite{zheng2022alpa,li2022amp,unger2022unity,miao2022galvatron} do not consider communication overlapping. 
Hence, another primary objective of this paper is to \textbf{automatically find the model parameter partition strategies for overlapped TMP training schedules}.

To address the above issues, we propose Oases in this paper, an operation overlapping and automated model partitioning approach to accelerate TMP training of large-scale models on commodity servers. 

Oases features two novel modules. The first is called \textit{fine-grained overlapping TMP training schedule}, which maximizes overlaps communication with computation, thereby accelerating TMP training. 
In this module, we decouple the data dependency between operations in TMP by data splitting, enabling operation scheduling in a sub-batch granularity. 
Moreover, we develop a \textit{cross-pass overlapping schedule} to break the barriers between passes, which allows us to schedule operations without interruption and pursue a better communication-computation overlapping.  
We also propose a \textit{fine-grained recomputation strategy} that determines which intermediate tensors to retain or discard during the forward pass, based on the observed redundancy in communications of recomputation pass.
This enables the communication reduction in the recomputation and more computation operations for overlapping.

Another module in Oases is named \textit{Oases planner}, which is an automatic model parallel strategy planner to attain further accelerations on TMP training. 
This planning involves devising distinct model parameter partition strategies for each partitioned operator in TMP to ensure better utilization of training resources.
In this module, we propose a \textit{new cost model} which can estimate: i) the performance of operations considering the training schedules with overlapped communications, and ii) the resharding overhead from adjacent operations with different parallel strategies. 
We then transform the parallel strategy optimization process into a cost minimization problem, constrained by the memory of devices. The cost data of a commodity cluster is obtained through offline profiling. Finally, this problem can be formulated as an integer linear programming (ILP) problem, effectively solvable using existing solvers.

The contributions of Oases can be summarized as follows:
\begin{enumerate}
\item[$\bullet$]  We present a fine-grained overlapping schedule of Oases, which jointly schedules the forward, recomputation, and backward passes of TMP training. Additionally, Oases can avoid communication in recomputation by utilizing its redundancy. The data-dependent communication operations of TMP can be overlapped with a maximum number of computation operations.

\item[$\bullet$]  We design the Oases planner which can model the performance of overlapped TMP training schedules and efficiently search for a parallel strategy with the best performance. To the best of our knowledge, Oases is the first work to model communication-computation overlap for TMP parallel strategy searching.

\item[$\bullet$]  We comprehensively evaluate the performance of Oases with models of varying sizes on two commodity GPU clusters. Compared to four popular model-parallel training methods including the state-of-the-art libraries, Oases accelerates the overall training process with up to 1.48\(\times\) speedup over the best baseline, and up to 1.95\(\times\) speedup and 2.18\(\times\) device utilization over Megatron. Our implementation is open-sourced at \url{https://github.com/lucasleesw/oases}.

\end{enumerate}
\section{Background and Motivation}

\begin{figure}
  \centering  
  \begin{subfloat}[The training operations of a \(L\) transformer layer model with TMP.]{\includegraphics[width=1\linewidth]{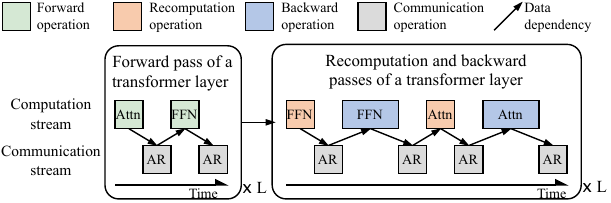}
\label{fig:sche_1}}
  \end{subfloat}

\begin{subfloat}[The model weight distribution in a transformer layer.]{
  \includegraphics[width=0.9\linewidth]{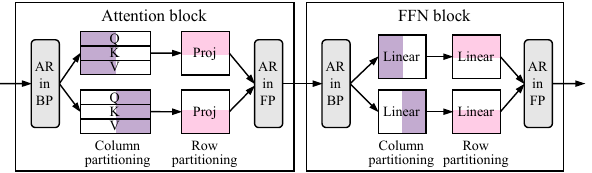}
  \label{fig:tp_bg_2}}
  \end{subfloat}
  \caption{Tensor model parallelism (TMP) in commodity servers, where Attn denotes the attention block, FFN denotes the feed-forward network, and AR denotes the AllReduce primitive.}
  \label{fig:tp}
\end{figure}

\noindent\textbf{Large-scale model architecture.}
Most recent large-scale models~\cite{brown2020language,touvron2023llama,zhang2022opt,scao2022bloom} are based on Transformer~\cite{vaswani2017attention} architecture. Researchers scale models to various sizes by stacking transformer layers with different hidden sizes. 
An example computation schedule of a transformer decoder model with \(L\) layers is shown in Fig.~\ref{fig:tp}. 
Each transformer layer comprises an attention block and a feed-forward network (FFN). 
In each training iteration, each layer first performs forward passes in order and then backward passes in reverse order. Backward passes typically have twice the computation overhead of forward passes.
The heavy computational loads of transformer blocks are matmul, and their parallelization has been well studied~\cite{georganas2012communication,van1997summa}.

\noindent\textbf{Tensor model parallelism (TMP).} 
TMP is a model-parallel approach proposed by Megatron ~\cite{shoeybi2019megatron} and is well-known for large-scale transformer model training~\cite{jia2022whale,lai2023merak,ao2021end,li2023colossal}.
TMP is an expert-optimized strategy for transformer-based models as shown in Fig.~\ref{fig:tp}. 
It partitions the weight matrices along the row dimension (requiring AllReduce in forward pass) or along the column dimension (requiring AllReduce in backward pass).
Although TMP can reduce memory usage in model training, it introduces many data-dependent AllReduce communications during both forward and backward passes, slowing down the overall training process.

\noindent\textbf{Recomputation.}
The intermediate results during the forward pass are called activations, which are used by the backward pass for gradient calculation but demand a large amount of memory. Shown in Fig.~\ref{fig:tp}, the recomputation~\cite{jain2020checkmate, chen2016training,kirisame2021dynamic} evicts activations and recomputes them when necessary. This approach can significantly reduce memory requirements, preserving more model parameters on single device, and is crucial for large-scale model training in memory-limited commodity servers.
Recomputation is widely adopted~\cite{liang2023survey,ren2023pangu,jia2022whale,smith2022using} in large-scale model training, but its default implementation~\cite{paszke2019pytorch} simply runs the forward function again. 

\subsection{Motivational study}

To reduce the communication overhead of TMP,
Wang et al.~\cite{wang2022overlap} and AccTFM~\cite{zeng2022acctfm} propose intra-op overlapping methods where they decouple and overlap the communication and computation operations of matrix productions. Merak~\cite{lai2023merak} suggests an inter-op overlapping schedule by splitting batches within forward and backward passes, but it fails to incorporate the recomputation technique, which is also necessary for giant model training on commodity hardware. 

\noindent\textbf{Case study.}
Fig.~\ref{fig:motivation} shows breakdowns of TMP training iterations in two transformer-based models using the popular Megatron~\cite{shoeybi2019megatron} framework on four 3090 GPUs. 
Communication comes from TMP accounts for 64.7\% and 59.2\% of overall training time in Megatron, which becomes a significant factor in degrading training performance and wasting resources.
As shown by the Oases bar, communication in Oases contributes much less to each iteration. \textit{A large amount of communication is overlapped through joint scheduling} of forward, recomputation, and backward operations, with the help of Oases fine-gained training schedule (Section~\ref{sec:schedule}).
And \textit{the overall iteration time of Oases is less than the communication time of Megatron}, as Oases estimates the redundant communication (Section~\ref{sec:schedule_2}) during recomputation and reduces the communication volume by searching for the best model partitioning scheme with Oases planner (Section~\ref{sec:planner}).

\begin{figure}[tb]
  \centering
  \includegraphics[width=0.9\linewidth]{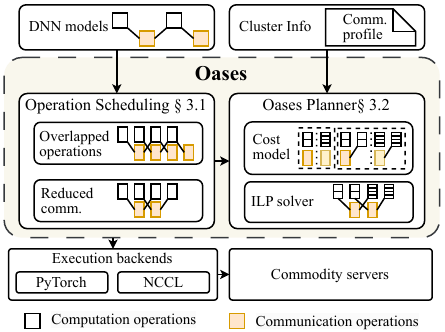}
  \caption{Oases overview.}
  \label{fig:overview}
\end{figure}

\section{Oases Overview}

As shown in Fig.~\ref{fig:overview}, Oases features two novel modules.
The first module is a \textbf{fine-grained overlapping TMP operation schedule}, which maximizes overlaps communication with computation, thereby accelerating TMP training on servers. In this module,
we develop a \textit{cross-pass overlapping schedule} to break the barriers between passes, which allows us to schedule operations without interruption and pursue a better communication-computation overlapping.  
We also propose a \textit{fine-grained recomputation strategy} to arrange the tensors saved for recomputation during forward, enabling the communication reduction in the recomputation and more computation operations for overlapping.

The second module is an automatic model parallel strategy planner named \textbf{Oases planner}, to attain further accelerations.
In this module, we propose a \textit{new cost model} which can estimate: i) the performance of operations considering the training schedules with overlapped communications, and ii) the resharding overhead from adjacent operations with different parallel strategies. 
Then, we formulate the parallel strategy optimization process under the memory footprint constraint as an integer linear programming (ILP) problem, which can be solved with existing solvers efficiently.

\begin{table} \renewcommand\arraystretch{1} 
  \caption{Major notations and their meanings in the training schedule}
  \label{tab:notation}
  \centering
  \begin{tabular}{cc}
    \toprule
    Notation     & Meaning \\
    \midrule
    \(F_i\) & Forward computation of the \(i^{th}\) operator   \\
    \(R_i\)     & Recomputation of the \(i^{th}\) operator  \\
    \(B_i\)    & Backward computation of the \(i^{th}\) operator       \\
     \multirow{2}*{\(C_i(\cdot)\)}   & \multirow{2}*{\shortstack{Communication of the \(i^{th}\) operator in forward (F), \\ recomputation (R), or backward (B)}}  \\
     & \\
    \bottomrule
  \end{tabular}
\end{table}

\section{Overlapping TMP training schedule}\label{sec:schedule}

In this section, we introduce the fine-grained overlapping training schedule of Oases, which schedules the forward pass, recomputation, and backward pass in a unified execution plan to accelerate the TMP training of large-scale models.
We first design the cross-pass schedule which breaks the barrier passes during backward, thereby scheduling both recomputation and backward. Next we schedule the operations in forward to drop the redundant communication, with the proposed fine-grained recomputation strategy.
Table~\ref{tab:notation} lists common notations used in this work.

\subsection{Cross-pass scheduling}\label{sec:schedule_1}

We can take the forward operator sequence \((F_i,C_i(F),F_j)\) as an example in TMP training, which is consistent with the FFN and attention layers in transformer-based models. And its recomputation enabled backward pass will be \((R_i,C_i(R),R_j)\),\((B_j,B_i,C_i(B))\). Fig.~\ref{fig:sche_1} illustrates the default backward schedule, the blocked operators result in low device utilization. 
This default implementation of recomputation involves dividing the continuous backward process into interleaved recomputation and backward passes, resulting in barriers between passes, as the recomputation pass directly aligns with the forward pass.

\begin{figure}
  \centering  
  \begin{subfloat}[The default TMP schedule. The communication and computation operations are blocked.]{\includegraphics[width=1\linewidth]{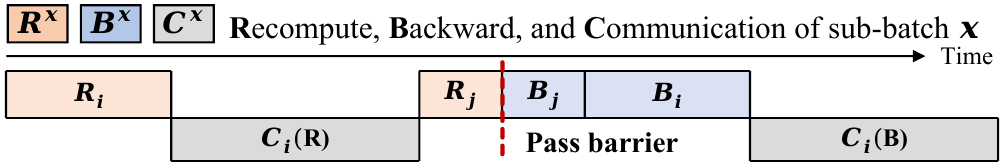}%
\label{fig:sche_1}}
  \end{subfloat}
  
  \begin{subfloat}[Scheduling the recomputation (inherited from forward) and backward pass individually. ]{
      \includegraphics[width=1\linewidth]{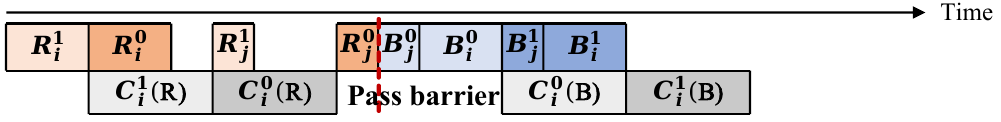}
      \label{fig:sche_2}}
  \end{subfloat}  
  
  \begin{subfloat}[Cross-pass overlapping schedule, which jointly schedules the recomputation and backward pass.]{
      \includegraphics[width=1\linewidth]{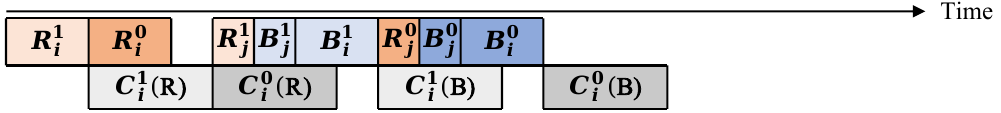}
      \label{fig:sche_3}}
  \end{subfloat}  
  
  \begin{subfloat}[Oases schedule. Recomputation communication is reduced with a fine-grained recomputation strategy.]{
      \includegraphics[width=1\linewidth]{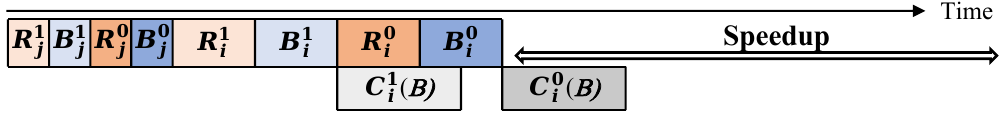}
      \label{fig:sche_4}}
  \end{subfloat}
  \caption{Execution timelines of computation (upper) stream and communication (lower) stream during TMP backward. Speedup is measured relative to the baseline Megatron training implementation, under the same hardware and configuration.}
  \label{fig:schedules}
\end{figure}

The computation and communication streams can work independently in most deep-learning systems. 
A straightforward solution is to partition the data batch into two sub-batches. We can use superscripts, such as \((R_i^0, R_i^1)\), to represent the related operator of sub-batches. 
Data dependency only relies on the operators with identical superscripts, the operations with different superscripts can be pipelined. 
Take the operator \(i\) in backward as an example, its communication of the first sub-batch and its backward computation of the second sub-batch, i.e., \(C_i^0(B)\) and \(B_i^1\), can be executed simultaneously and overlapped.
As shown in Fig.~\ref{fig:sche_2}, this basic pipelined backward schedule can speed up the recomputation and backward passes individually, making it practical in non-recomputation scenarios~\cite{lai2023merak}. 
However, the consistency of operations is restricted due to the pass barriers.

\noindent\textbf{Crossing the propagation pass barrier.}
To jointly optimize the recomputation and backward, 
we design \textbf{a cross-pass overlapping schedule.} 
In our implementation, we develop a backpropagation runtime engine, where we perform recomputation from scratch instead of invoking the forward function again, and we schedule the recomputation and backward operations in the sub-batch granularity. 
The scheduling rule lies in overlapping communication with computation operations as much as possible.
This approach allows us to break the barriers between passes in the backward, thereby overlapping communications with cross-pass computation operations.
The backpropagation order of sub-batches might be adjusted accordingly, while ensuring that the results are not affected.
As the example shown in Fig.~\ref{fig:sche_3}, we merge the operator sequence of each sub-batch and schedule them simultaneously in our cross-pass schedule, \((R_j^1, B_j^1, B_i^1)\) can be executed continuously and the communication of recomputation \(C_i^0(R)\) can be properly hidden. 
Generally the granularity of recomputation is the transformer layer~\cite{narayanan2021efficient}, and each of them will cause a pass barrier. Thus breaking these barriers will bring a remarkable acceleration.

\subsection{Redundant communication in recomputation}\label{sec:schedule_2}

As described in Section \ref{sec:schedule_1}, an FFN or attention layer will involve three communication operators in a TMP training iteration.
The communication primitive is AllReduce, 
we can formulate the output \(y\) of an AllReduce operation for given inputs \(x\) among \(w\) devices as \(y=\sum^w_{i=0}x_i\), where we can get \(\frac{\partial y}{\partial x_i} = 1\). Denoting the loss function as \(\phi\), we have:
\begin{equation}\label{eq:1}
    \frac{\partial \phi}{\partial x_i} = \frac{\partial \phi}{\partial y} \frac{\partial y}{\partial x_i} = \frac{\partial \phi}{\partial y} .
\end{equation}
Equation~\eqref{eq:1} implies that the gradients of AllReduce's inputs coincide with the gradients of its outputs. In other words, when the AllReduce serves as the output of the recomputation sequence, it can be omitted during the recomputation. Thus we can propose \textbf{a fine-grained recomputation strategy}, which initializes recomputation after communication operators in forward. 
Consider the example forward operators \((F_i^1,C_i^1(F),F_j^1)\), we can apply recomputation on both \((F_i^1,C_i^1(F))\) and \((F_j^1)\), its backward will become \((R_j^1, B_j^1, R_i^1, B_i^1, C_i^1(B))\), the communication in recomputation \(C_i^1(R)\) is eliminated. 
Fig.~\ref{fig:sche_4} shows the backward schedule with fine-grained recomputation, the number of communication is reduced, and the communication of backward can be overlapped with more recomputation operations. 

When training with multiple FFN and attention layers, our backward schedule can run across layers and the speedup can be more significant. 
Since we only change the start positions of each recomputation sequence, the number of sequences is not increased and we will not store extra input data.

\subsection{Overlapped operation runtime}
To realize the fine-grained overlapping training schedule in Oases, we implement a new runtime executor for operation scheduling. 
The deep neural network operators can be classified into two types: computation operators and communication operators. 
Each computation operator needs three operations in order: \texttt{Forward}, \texttt{Recompute}, and \texttt{Backward}. 
The \texttt{Recompute} operations can be collectively executed within the operator sequence, enabling us to only store the input data of the first operator in the sequence of recomputation.
For communication operators, they can be executed asynchronously following a \texttt{StartAsync} instruction and continue as such until a \texttt{Sync} instruction is made.

Now we can summarize our training schedule in Algorithm~\ref{alg:schedule}.
Deep neural network training can be presented by a directed acyclic graph. 
We can flatten the graph to an operator sequence \(O\). 
In the forward schedule, we split the data batch into two sub-batches (line 3) and execute their related operators as follows: i) we initiate \texttt{Forward} operations for computation operators;  
and ii) for communication operators, we start them asynchronously and switch the sub-batch, through the \texttt{Sync} and \texttt{StartAsync} instructions (lines 6--13). 
Specifically, we save the operators which have finished \texttt{Forward} operation into a sequence after each communication operator for recomputation. And we drop the communication operations in this stored sequence (line 11). 
In the backward schedule, we can straightforwardly follow the saved operator sequence from the forward in reversed order. We perform \texttt{Recompute} operations on the saved operators to get the corresponding backward operator sequences and perform \texttt{Backward} operations (lines 17--21). 
The communications in backward operator sequences will be overlapped by computation including both \texttt{Recompute} and \texttt{Backward} operations (lines 23--27).
Some operations in the warm-up and cool-down phases are omitted for simplification. 
With our training schedule, operations from different sub-batch form a pipeline: when one sub-batch is computing, another sub-batch will communicate.

\begin{algorithm}[tb]
  \KwIn{Model operator sequence \(O\)}
  \SetKwProg{forward}{At Forward}{:}{}
  \forward{}{
    \(x,i,j \gets 0,  handler \gets null\)\\
    \((o_0^0, o_0^1,...,o_k^0, o_k^1) \gets \) Split(\(O\))\\
    \(A\gets[\ ]\)  \tcp{Init the operator sequence array for recomputation}
    \While{\( i \leq k\)}
    {
        \eIf{\(o_i^x\) is a computation operator}
        {   
            Forward(\(o_i^x\)), \(i \gets i+1\) \\
        }
        {
            \tcp{Synchronize previous communication and start asynchronous communication}
            Sync(\(handler\)) \\
            \(handler \gets \)StartAsync(\(o_i^x\)) \\
            \tcp{Save operator sequence for recomputation}
            \(A.append((o_j^x,...,o_{i-1}^x)) \) \\
            \lIf{\(x=1\)}{
            \(j \gets i+1 \)}
            \(x \gets x~\text{XOR}~1, i\gets j\) 
        }
    }
    } 
  \SetKwProg{backward}{At Backward}{:}{} 
  \backward{}{
    \(handler \gets null, O_b' \gets [\ ]\) \\
    \While{\(A\) is not empty}{
      \(O \gets A.PopLast \) \\
      \tcp{Get backward operators by recomputing}
      \(O_b \gets \)Recompute(\(O\)) \\
      \While{\(o \gets O_b.PopLast \neq null \) }
      {
        \eIf{\(o\) is a computation operator}{
            Backward(\(o\)) \\
        }{
            Sync(\(handler\))\\
            \(handler \gets \)StartAsync(\(o\)) \\
            \tcp{Finish interrupted backward operations}
            \While{\(o' \gets O_b'.PopLast \neq null \)}
                {Backward(\(o'\)) \\
                }
            \(O_b' \gets O_b\) \\
            \textbf{Break}
        }
      }
    }
    } 
\caption{Oases training schedule.}
\label{alg:schedule}
\end{algorithm}

\begin{figure}
\centering
\includegraphics[width=0.65\linewidth]{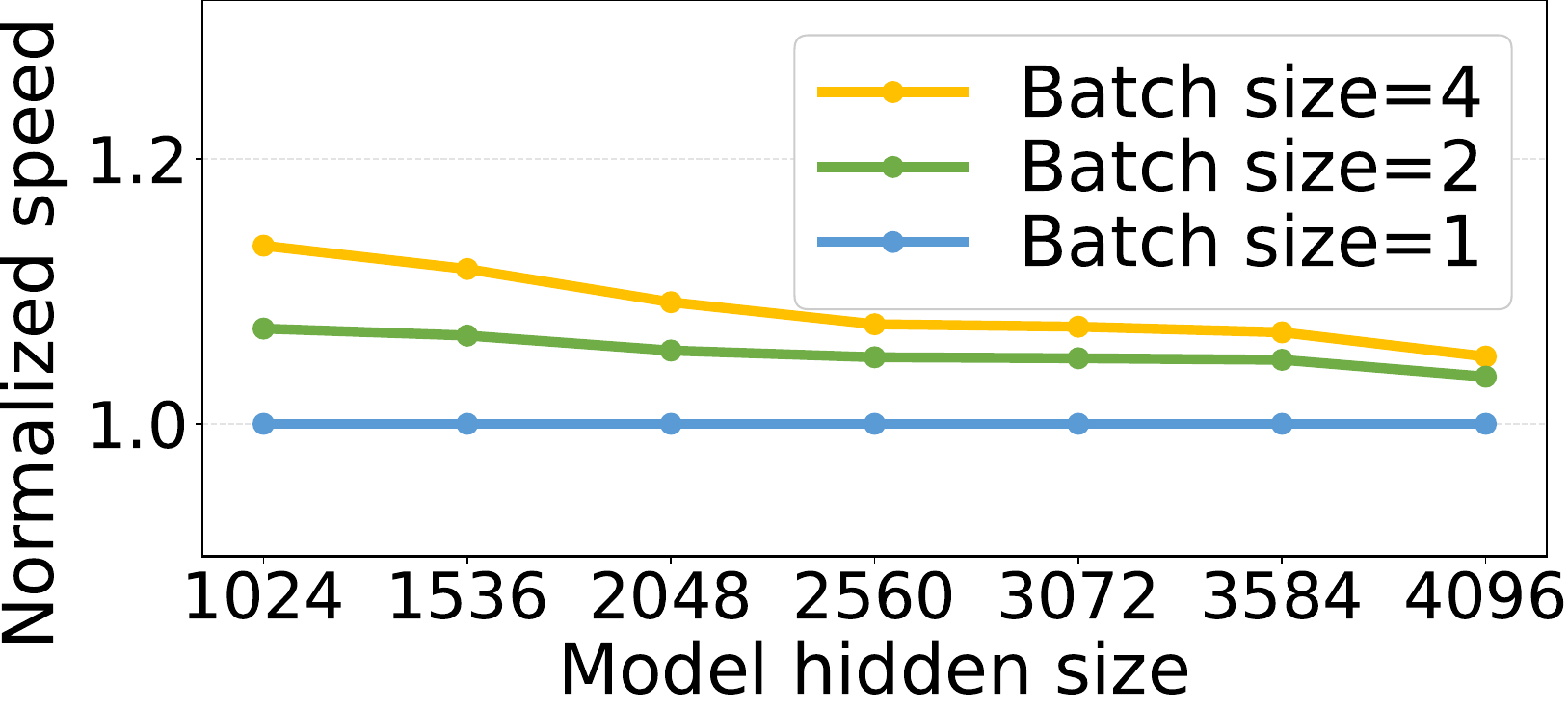}
\caption{Normalized training throughput for varying batch sizes across models with different hidden sizes.}
\label{fig:batch_hidden}
\end{figure}

The overlapped training schedule in Oases requires batch splitting, which may reduce arithmetic density. To evaluate this effect, we measure training performance across different batch sizes on models with varying hidden sizes. The results, shown in Figure~\ref{fig:batch_hidden}, indicate that the performance loss due to reduced batch size is less than 20\%, and tends to be smaller for models with larger hidden sizes. We consider this loss in arithmetic density to be an acceptable trade-off given the overall performance gains achieved through overlap.

\section{Parallel strategy planner} \label{sec:planner}

With our fine-grained overlapping schedule in Section~\ref{sec:schedule}, the communication of TMP is overlapped by model computation as much as possible. 
We have two observations for further acceleration:
i) The parallel strategy of TMP is uniform across model layers, but there is a considerable memory consumption gap between different TMP parallel strategies. 
We can model the communication volume of an AllReduce~\cite{patarasuk2009bandwidth} as \(2K(N-1)/N\), where \(K\) is the message size and \(N\) is the TMP degree. This indicates a TMP module can trade extra memories for smaller \(N\) and thus less communication volume. 
And ii) in our schedule, the communication overhead during forward is equal to backward, while the computation overhead in the forward is approximately 1/3 of it in the backward (includes recomputation overhead). The communication in the forward may be exposed and cause performance bottlenecks. 

Based on the above observations, finding a proper parallel strategy of TMP for large-scale models has potential performance benefits.
But it is challenging to model the complicated overlapping training schedule. To address the problem, we propose the \textbf{Oases planner}, which can automatically generate a high-throughput parallel training strategy.

\subsection{Problem formulation}

\noindent\textbf{Searching space.} Given an model graph \(G\), global batch size \(b\), device memory capacity \(m\), and possible parallel strategies \(p\) of a training cluster, 
the goal of Oases planner is to maximize the training throughput.
We assume the model runs with a uniform TMP strategy to initialize communication operators.
The parallel training strategy \(s\), inspired by Alpa~\cite{zheng2022alpa}, can be represented by a collection of one-hot decision vectors \(s_i \in \{0, 1\}^p\subseteq \mathbb{R}^{p}\) for each graph node \(i\). \(s_{ij} = 1\) means node \(i\) is using the \(j\)-th partitioning scheme.
Let \(S\) denotes the entire space of strategies and \(f(\cdot)\) estimates per iteration execution time for strategy \(s\). 
We can formulate the optimization problem of Oases planner as follows:
\begin{equation}\label{eq:argmin}
s^*=\mathop{\arg\min}_{s\in S} ~ f(s; G,b,m,p) .
\end{equation}

\noindent\textbf{Simplification.} To simplify the model graph under our fine-grained overlapping TMP training schedule, we merge computation operators between adjacent communication operators into computation sequences. 
The resulting forward and backward model graphs are composed of alternating computation sequences and communication operators.
Model graphs typically start with a computation sequence, we can further group each computation sequence with its subsequent communication operator, forming a block that comprises a computation sequence and a communication operator. 
Consequently, we can update the model graph as \(G(V, E)\), where each node \(v \in V\) is an 
aforementioned block and \(E\) is the set of edges that connect pairs of nodes.

\subsection{Cost model}\label{sec:costmodel}

\noindent\textbf{Cost of overlapped operations.}
We can estimate the \(f(\cdot)\) in \eqref{eq:argmin} by summing the costs of all nodes and edges, conditioned on the user-specified model, global batch size, and training cluster. 
For each node \(i\), it has computation cost vectors \(d_i \in \mathbb{R}^p \) and communication cost vectors \(c_i \in \mathbb{R}^p \). The operation costs of a given parallel strategy can be presented by the dot production, e.g., \(s_i^{\top} c_i\).
Additionally, we denote  \(d(F), d(B)\) and \(c(F), c(B)\) as the corresponding forward and backward cost vectors, and \((\cdot)^{\top}\) denotes the transpose of a vector. 
For the forward graph nodes, as described in Algorithm~\ref{alg:schedule}, each block is executed with two sub-batches. 
Computation and communication can be aligned by right-shifting the communication operation of sub-batches. Specifically, the computation of a sub-batch will overlap the last communication of another sub-batch, and the cost after overlapping can be estimated using the $\max$ function.
 Therefore, we can present the cost of node \(i\) by summing the computation costs of two sub-batches, i.e., \(\mathop{\max}\{s_i^{\top}  d^0_i, s_{i-1}^{\top}  c^1_{i-1}\}  + \mathop{\max}\{ s_i^{\top}  d^1_i, s_i^{\top}  c_i^0\}\). The node costs in backward can be estimated in the same method with different cost vectors. When considering the first and last node, the node cost \(T_V\) of \(k\) nodes is formulated as follows:
\begin{equation}\label{eq:node}
\begin{aligned} 
T_V(d,c) = & s_0^{\top}  d_0^0   + \sum_{i=1}^{k} \mathop{\max}\{s_i^{\top}  d^0_i, s_{i-1}^{\top}  c^1_{i-1}\} + \\
& \sum_{i=0}^{k} \mathop{\max}\{ s_i^{\top}  d^1_i, s_i^{\top}  c_i^0\}+s_k^{\top} c_k^1 .
\end{aligned}
\end{equation}

\noindent\textbf{Resharding cost.}
When using different parallel strategies between adjacent nodes, an blocked AllGather communication is required for data resharding, converting the tensor layouts between nodes. 
These additional costs can be treated as edge costs of the given model graph \(G(V,E)\), which include two aspects: i) the communication overhead of the AllGather \(T_\mathrm{AG}\), and ii) the costs of adjacent nodes that can no longer overlap because of the blocked AllGather operation, which can be calculated with the $\min$ function. 
Let \(\mathcal{R}_{vu} \in \mathbb{R}^{p\times p}\) denotes the edge cost matrix, where the element \(\mathcal{R}_{vuij}\) is the edge cost from the output of \(i\)-th strategy of node \(v\) to the input of \(j\)-th strategy of node \(u\). 
The AllGather happens during the forward when \(i<j\), and during the backward when \(i>j\). 
To this end, we can formulate each edge cost as follows:
\begin{equation}\label{eq:edge}
 \mathcal{R}_{vuij}=
\begin{dcases}
T_\mathrm{AG}(F,i,j) + \mathop{\min}\{ s_v^{\top}  c_v^1(F) , s_u^{\top}  d_u^0(F)\}, & \text{ \(i < j\)} \\
T_\mathrm{AG}(B,i,j) + \mathop{\min}\{s_v^{\top}  c_v^1(B) , s_u^{\top}  d_u^0(B)\}, &\text{ \(i > j\)} \\
0, & \text{ \(i = j \)}
\end{dcases}
\end{equation}

\begin{table*} 
\caption{The example parallel strategies (TMP degrees for each transformer layer), the optimization time of Oases planner, and the corresponding runtime performance.}
\label{tab:scheme}
\centering
\begin{tabular}{ccccccccc}
\toprule
 \multirow{2}{*}{\shortstack{Model\\layer name}} & \multirow{2}{*}{\shortstack{\# of\\layers}} & \multirow{2}{*}{Algorithm} & \multirow{2}{*}{\shortstack{Parallel strategy \\ (TMP degree$\times$\# layers)}} & \multirow{2}{*}{\shortstack{Optimization time\\per run (ms)}} & \multirow{2}{*}{\shortstack{Reshard cost\\per iteration (ms)}} & \multirow{2}{*}{\shortstack{TMP cost\\per iteration (ms)}} & \multirow{2}{*}{\shortstack{Throughput\\(k tokens/s)}} \\
  &  &  &  &  &  \\
\midrule 
 \multirow{2}{*}{BERT-1.3B} & \multirow{2}{*}{24} & w/o Planner & {[}{[}4{]}$\times$24{]} & - &  - & 374.8 & 43.9 \\
   &  & w/ Planner & {[}{[}2{]}$\times$8+{[}4{]}$\times$16{]} & 451.6 &  17.6 & 241.9 & \textbf{50.9} \\
\midrule 
 \multirow{2}{*}{LLaMA-7B} & \multirow{2}{*}{16} & w/o Planner & {[}{[}4{]}$\times$16{]}  & - & -  & 103.1 & 23.8 \\
  &  & w/ Planner & {[}{[}2{]}$\times$10+{[}4{]}$\times$6{]} & 380.4 &  6.5  & 42.2& \textbf{25.2} \\
\midrule  
  \multirow{2}{*}{LLaMA-65B} & \multirow{2}{*}{8} & w/o Planner & {[}{[}8{]}$\times$8{]} & - & - & 187.5 & 8.9 \\
   &  & w/ Planner & {[}{[}4{]}$\times$2+{[}8{]}$\times$6{]} & 157.7 & 12.1  & 124.0 & \textbf{9.1} \\
 \bottomrule
\end{tabular}
\end{table*}

\subsection{Finding the best strategy}\label{sec:ILP}

Combining the \eqref{eq:node}--\eqref{eq:edge}, we can express \(f(\cdot)\) in~\eqref{eq:argmin} by summing the node costs in forward, node costs in backward, and edge costs. And the cost minimization problem in~\eqref{eq:argmin} can be formulated as an ILP problem with the following objective:
\begin{equation}\label{eq:ilp}
    \mathop{\min}_s ~ \Big\{T_V(d(F),c(F))+ T_V(d(B),c(B)) + \sum\limits_{(v,u)\in E}s_v^{\top}  \mathcal{R}_{vu}  s_u \Big\} .
\end{equation}

\noindent\textbf{Memory constraints.}
To ensure the successful execution of the parallel strategy without memory issues, the optimization should take into account the limitations of the device's memory.
The memory consumption in TMP training includes three main aspects: (i) model states, which consist of the model weights and optimizer states; (ii) saved input tensors, which are saved during forward pass for recomputation (line 11 of Algorithm~\ref{alg:schedule}); and (iii) the temporary buffers for computation kernels, which is generally reused among operations. The memory consumption is related to the TMP degree.
Similarly, we denote the memory cost vectors of model states, saved input tensors, and temporary runtime buffers with \(m_s, m_t, m_r \in \mathbb{R}^p\), respectively.
We can estimate the memory consumption and derive the following ILP constraint of device memory: 
\begin{equation}\label{eq:mem}
   s_k^{\top}  m_r + \sum\limits_{v\in V} (s_v^{\top}  m_s + s_v^{\top}  m_t ) < m  .
\end{equation}

We can run offline profiling to get the accurate values for \(d(\cdot)\), \(c(\cdot)\), and \(m_r\) in the ILP formulation.
Recent transformer-based models feature a uniform and repeating architecture, allowing profiling to be performed on a single transformer layer. During profiling, this layer is executed under all possible TMP parallel strategies while dedicated timers record the associated costs. Each strategy takes approximately 40 seconds to profile. To reduce complexity, we limit the possible partitioning schemes to the power of two. For example, training a large-scale model on 32 GPUs involves 6 possible TMP strategies, resulting in a total profiling time of about 4 minutes. 
This one-time cost can be amortized across multiple training runs and only needs to be repeated if the cluster configuration or workload significantly changes.
We model the overhead of AllGather~\cite{thakur2005optimization} with message sizes, and count \(m_s\) and \(m_t\) according to the model config and our training schedule.   
Now we can solve \eqref{eq:ilp}--\eqref{eq:mem} optimally with an open source solver~\cite{forrest2005cbc}. 

Table~\ref{tab:scheme} presents example parallel strategies (TMP degrees) for each transformer layer after applying the Oases planner and the time taken for Planner optimization, along with the runtime performance.
A decreasing TMP degree of some layers will accelerate the model training, as it reduces the TMP communication at a small reshard cost per training iteration.
Although the Planner overhead increases with the number of model layers, all optimization processes are completed within only half a second in all our experiments.

In summary, Oases leverages a fine-grained training operation schedule to maximize overlapping communication and computation in TMP training. 
The Oases planner further identifies the best model parallel strategy based on the overlapped training schedule.
Oases is generally applicable to accelerate TMP training, particularly in training scenarios with limited communication bandwidth between TMP workers.
For instance, it can be effectively applied for TMP training large-scale models on commodity GPU servers, provided that high-speed NVLinks are absent.

\begin{table} \renewcommand\arraystretch{1} 
\small  
\caption{The model configurations of performance experiments. Note that we use fewer transformer layers than the original models.}
\label{tab:allconfig}
\centering
\resizebox{\linewidth}{!}{
\begin{tabular}{cccccc}
\toprule
\multirow{2}{*}{\shortstack{Model layer\\name}} &
\multirow{2}{*}{\shortstack{Model hidden\\size}} & \multirow{2}{*}{\shortstack{Number of\\used layers}} &  \multirow{2}{*}{\shortstack{TMP\\degree}} & \multirow{2}{*}{\shortstack{Global\\ batch size}} \\
 &  &  &  &  &  \\
\midrule
BERT-large~\cite{devlin2018bert}&1024 & 24  & 2  & 256 \\
BERT-1.3B~\cite{shoeybi2019megatron}&2048 & 24  & 4  & 128 \\
GPT-8.3B~\cite{shoeybi2019megatron}&3072 & 24  & 4  & 32 \\
LLaMA-7B~\cite{touvron2023llama}&4096 & 16  & 4  & 32 \\
GPTNeoX~\cite{black2022gpt}&6144 & 16  & 8  & 8 \\
LLaMA-65B~\cite{touvron2023llama}&8192 & 8  & 8  & 8 \\
GPT-3~\cite{brown2020language}&12288 & 4  & 8  & 8\\
\bottomrule
\end{tabular}}
\end{table}

\section{Evaluations}

\subsection{Experiment setup}
\noindent\textbf{Cluster configurations.}
We evaluate Oases on two 8-node clusters with two types of commodity GPUs. 
The first cluster is \textbf{NVLink 3090}, where each node features four NVIDIA RTX3090 GPUs (24GB each) and four-slot NVLink 3.0~\cite{li2019evaluating} connections between GPU 0 and 1, and between GPU 2 and 3.
And nodes are equipped with four NVIDIA RTX3090 GPUs without NVLink in the second cluster \textbf{3090}.
All nodes are interconnected via 100 Gbps InfiniBand, and each GPU connects to the CPU via PCIe 4.0. 
All servers run 64-bit Ubuntu 18.04, CUDA 11.3, cuDNN 8.2.1, NCCL 2.10.3, GPU driver 510.68.02, and PyTorch 1.10.0.

\begin{figure*}[tb]
  \centering
  \begin{subfloat}[Performance on cluster NVLink 3090.]{
      \includegraphics[width=0.485\linewidth]{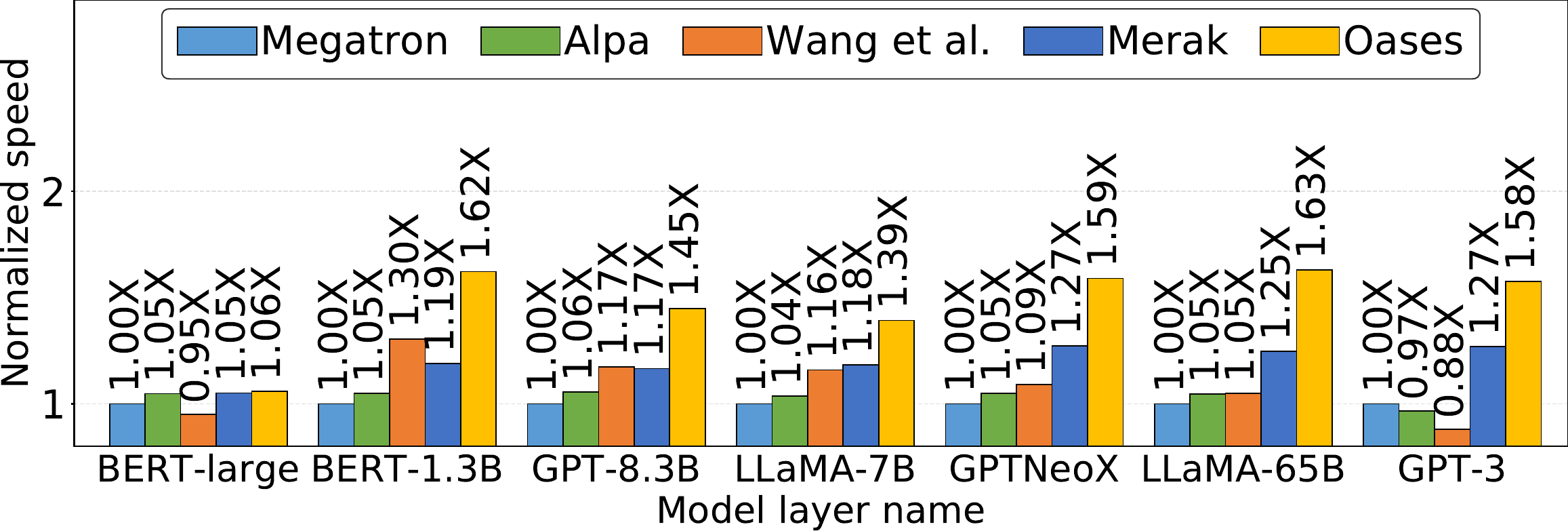}
      \label{fig:throught_1}}
  \end{subfloat}
  \hfill
  \begin{subfloat}[Performance on cluster 3090.]{\includegraphics[width=0.485\linewidth]{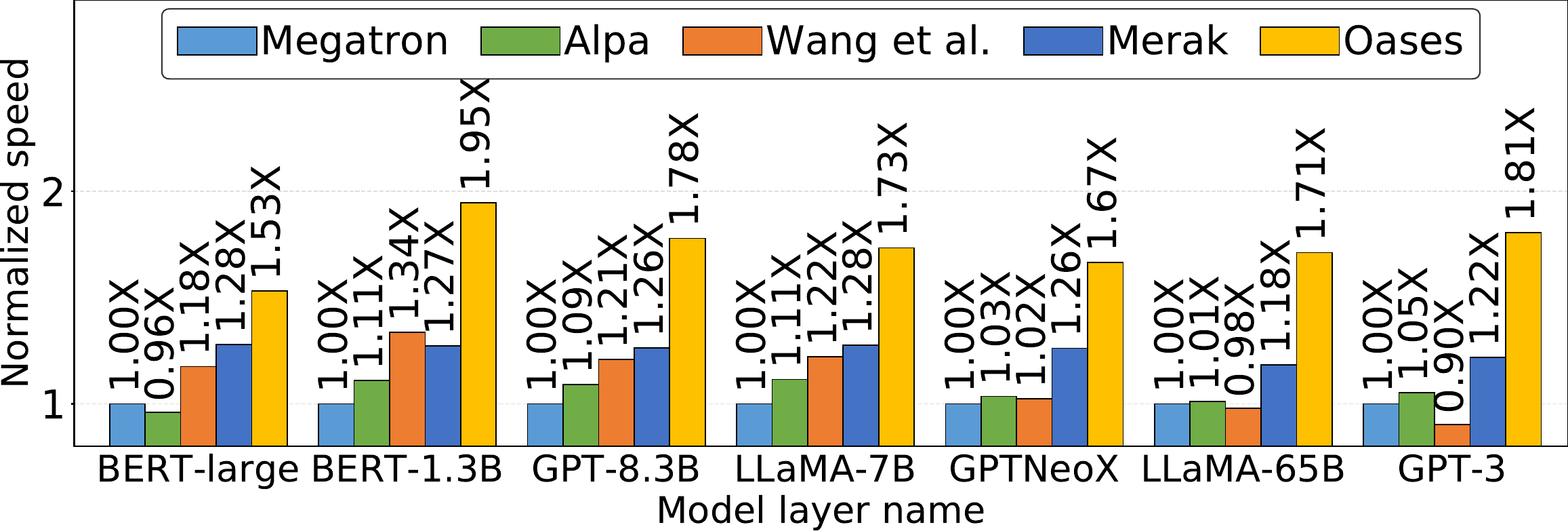}
      \label{fig:throught_2}}
  \end{subfloat}  
  \caption{End-to-end training performance on different model layers. Throughputs are normalized by Megatron.}
  \label{fig:throught}
\end{figure*}

\begin{table*}[tb] 
\small
  \caption{Averaged device utilization during TMP training of models in end-to-end experiments.}
  \label{tab:efficiency}
  \centering
  \begin{tabular}{ccccccccc}
    \toprule
    \multirow{2}*{Cluster} &\multirow{2}*{Methods}  & \multicolumn{7}{c}{Model layer name}\\
    \cmidrule(r){3-9}
     &  &BERT-large&BERT-1.3B&GPT-8.3B&LLaMA-7B&GPTNeoX&LLaMA-65B&GPT-3\\
    \midrule
        \multirow{2}*{\shortstack{NVLink\\3090}}  
    &  Megatron
    & 83.9\%
    & 47.2\%
    & 53.7\%
    & 57.5\%
    & 32.9\%
    & 37.4\%
    & 44.4\%  \\
    & Oases 
    & 97.8\%
    & 88.5\%
    & 90.1\%
    & 93.2\%
    & 62.7\%
    & 70.4\%
    & 77.9\% \\
    \midrule
    \multirow{2}*{\shortstack{3090}}  
    & Megatron
    & 56.2\%
    & 36.4\%
    & 43.2\%
    & 47.4\%
    & 28.6\%
    & 32.5\%
    & 39.7\% \\
    & Oases 
    & 96.4\%
    & 78.9\%
    & 87.0\%
    & 90.0\%
    & 62.3\%
    & 70.0\%
    & 77.9\% \\
    \bottomrule
  \end{tabular}
\end{table*}

\noindent\textbf{Models.}
Table~\ref{tab:allconfig} lists the model settings in experiments unless otherwise stated. 
The model settings are derived from previous studies~\cite{narayanan2021efficient,shoeybi2019megatron}, and are consistent with popular large-scale models.
The TMP degrees are the best strategies used in baselines and might be changed with Oases planner. 
It is worth noting that limited by device numbers, the numbers of transformer layers used in experiments are much smaller than general large-scale models. 
We believe evaluating some model layers is more substantial since TMP is usually used accompanied by PMP~\cite{huang2019gpipe,li2022harmony}, which is responsible for deepening models. In the experiment results, we denote the evaluated models with the model layer name.

\noindent\textbf{Baselines.}
We compare Oases with four popular approaches in experiments: 
(i) \textbf{Megatron}~\cite{narayanan2021efficient,shoeybi2019megatron,korthikanti2023reducing} proposes the TMP training for transformer-based models.
To ensure a fair comparison, we grid-search the TMP degree settings on a given model and global batch size, and report the best results.
(ii) \textbf{Alpa}~\cite{zheng2022alpa} is the current SOTA auto-parallelism training system, which can automatically generate an execution plan, but without overlapping communication with computation in TMP.
(iii) \textbf{Merak}~\cite{lai2023merak} is the SOTA 3D parallelism library for transformer-based models training. 
And (iv) \textbf{Wang et al.}~\cite{wang2022overlap} decouples communication and computation thus overlapping them in Einsum operators of models. Unfortunately it is close-sourced, we implement it based on Megatron to the best of our abilities.
We disable the PMP as it is orthogonal to our work unless otherwise stated.

\noindent\textbf{Metrics.} 
Oases is a synchronous method that can maintain the convergence of models, therefore we focus on comparing throughputs. 
Reported values are taken from an average of the latest 20 iterations~\cite{li2022amp}, out of a total of 100 training steps.

\subsection{End-to-end performance}\label{sec:evalend}

We compare the end-to-end performance of Oases with baselines on seven models and two commodity clusters. We report the normalized training throughputs in Fig.~\ref{fig:throught}.

Oases attains expected results for both NVLink 3090 and 3090 cluster, achieving speedups of 1.01--1.31\(\times\) and 1.20--1.48\(\times\)  over the best baseline, and up to 1.63\(\times\) and 1.95\(\times\) over Megatron, respectively.
The smallest speedup is observed for the BERT-large model on NVLink 3090, where the TMP degree is 2 and communications occur only through NVLink.
The communication performance bottleneck is not obvious under this high bandwidth scenario, as all methods exhibit similar results.
Alpa slightly outperforms Megatraon-LM in some cases with auto-searched parallel strategies,
but Megatron and Alpa present unsatisfactory speeds as they are unable to overlap communications of TMP. 
Though Merak partially overlaps communications and is the best baseline in most cases, Oases can offer more speedups through our fine-grained training schedule and optimized parallel strategy.
Wang et al.~\cite{wang2022overlap} performs well on smaller models (with hidden size \(\leq 4096\)). But for larger models with TMP degree 8, which requires inter-node communications, its decoupling increases the number of operations and slows down the training process.
Oases provides more remarkable speedups on cluster 3090, where communication is a more substantial performance bottleneck. 

Oases can find a more efficient partitioning scheme and effectively support overlapping communication with computation, especially during backpropagation, where the communication volume is reduced and communication could be overlapped with more computation. Thus, Oaese effectively accelerates TMP training on commodity GPU servers.

\subsection{Device utilization}\label{sec:evaldevice}
To study the reasons behind Oases' acceleration, we can consider device utilization. Exposed communications can idle devices, thereby reducing utilization. 
We apply stream multiprocessor efficiency~\cite{sanders2010cuda} of GPUs to represent the device utilization during training, and the results are shown in Table~\ref{tab:efficiency}. 
Our training schedule requires batch splitting, which potentially reduces arithmetic density. Therefore, some accelerations of the end-to-end throughput may not match the improvement of device utilization.
We think the resulting loss of arithmetic density is acceptable when compared to the gained performances. 
We observe that Megatron’s device utilization significantly drops when TMP communication is limited to NVLink (as in BERT-large), confined to intra-node links, or extended to inter-node communication (for models with hidden size $>$ 6144). This is because increased communication overhead, especially across inter-node boundaries, leads to greater idle time on GPUs, thus reducing overall utilization.

Compared to Megatron, Oases achieves 1.17--2.18\(\times\) higher device utilization, and its performance is less affected by bandwidth. These improved utilizations will result in overall training speedups, and make it possible to enable large-scale model training with TMP on commodity servers.

\begin{figure}[tb]
    \centering
    \begin{subfloat}[GPT-18.4B, Oases speedup Merak by 1.10--1.27\(\times\).]{
    \includegraphics[width=0.9\linewidth]{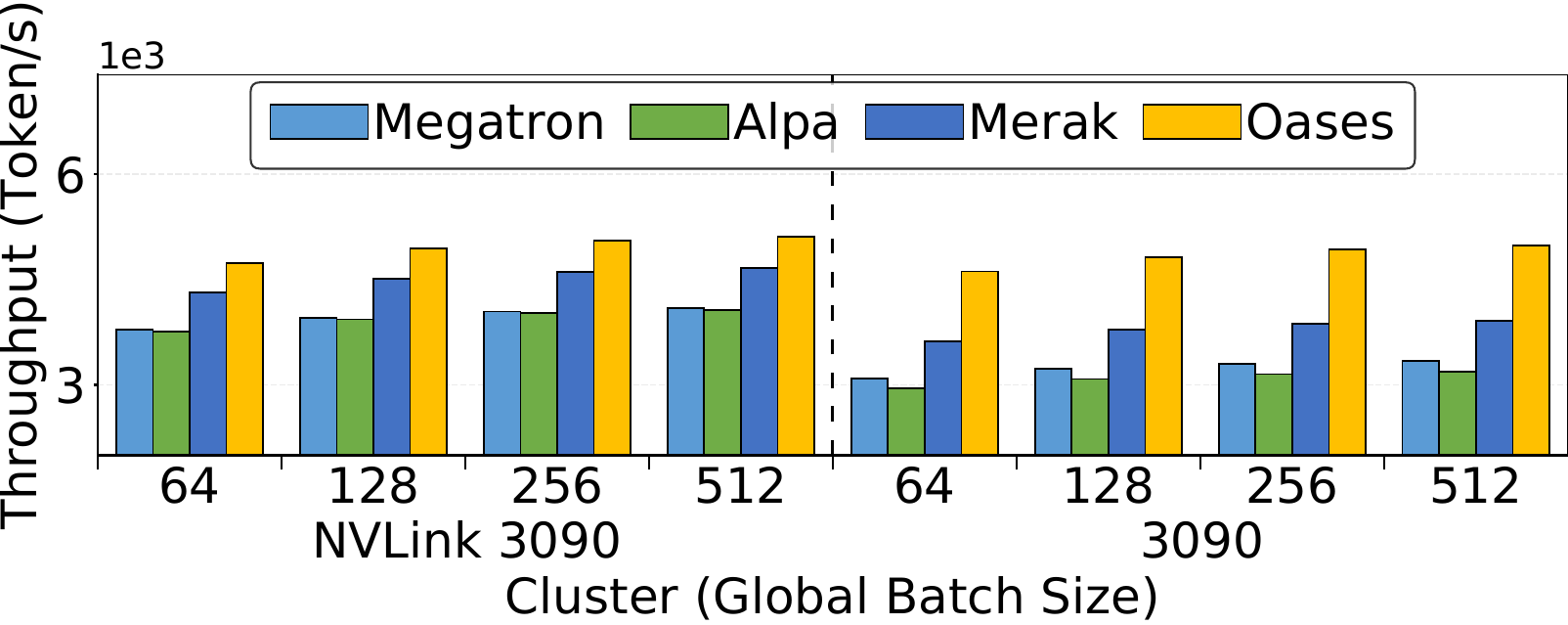}
    \label{fig:holderb}}
    \end{subfloat}
    \begin{subfloat}[GPT-39.1B, Oases speedup Merak by 1.31--1.35\(\times\).]{
    \includegraphics[width=0.9\linewidth]{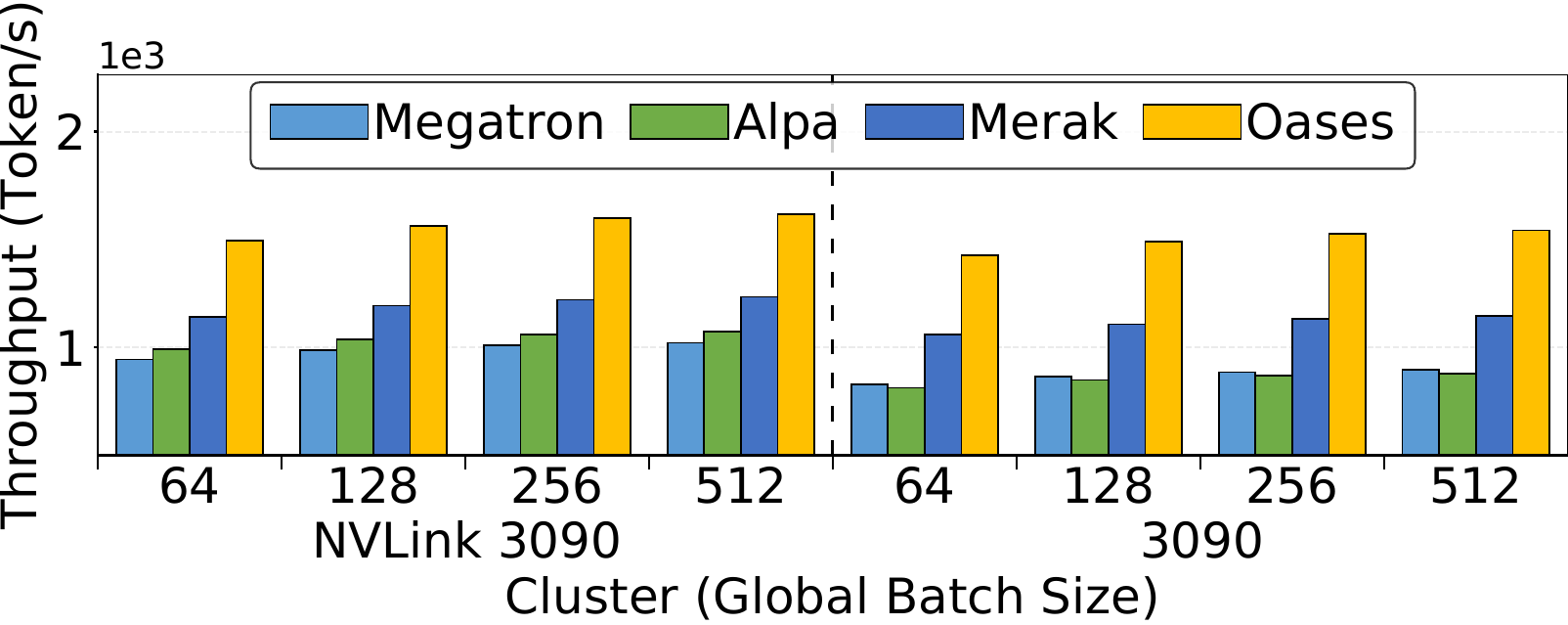}
    \label{fig:holdera}}
    \end{subfloat}
    \caption{Throughputs (tokens/s) of training complete models with pipeline model parallelism and Oases.}
    \label{fig:pmp}
\end{figure}

\subsection{Combining with pipeline model parallelism}\label{sec:evalpmp}
PMP is often combined with TMP to accommodate a complete model in large-scale model training. 
We integrate Oases into Megatron PMP and test it on GPT~\cite{brown2020language} models with parameter sizes of 18.4B and 39.1B. 
The model settings follow those of Megatron~\cite{narayanan2021efficient}.
To ensure an unbiased evaluation, we adopt the 1F1B PMP schedule~\cite{narayanan2021efficient}, and use the same PMP parallelism degree and micro batch size. 
The transformer layers are evenly divided into PMP groups. 
Fig.~\ref{fig:pmp} shows the throughput results of different global batch sizes. Oases integrated training improves speed by 1.10--1.35\(\times\) over the best baseline and 1.25--1.72\(\times\) over Megatron. 
For the GPT-18.4B model, Oases gets a modest improvement on cluster NVLink 3090.
This is because the fewer TMP degrees and fewer layers per device results in less communication in TMP, and improvements of TMP thus contributing less to the throughput.
When communication becomes a more apparent bottleneck on cluster 3090, the performance decline of Oases is more minor. And Oases achieves good acceleration on the GPT-39.1B model.

\begin{table*}[tb] 
\small
\caption{Ablation study. Throughput (k tokens/s) and speedups over Megatron. }
\label{tab:ablation}
\centering
\begin{tabular}{ccccccc}
\toprule
\small
 \multirow{4}*{Cluster} & \multirow{4}*{\shortstack{Model\\layer
 \\name}} & \multicolumn{5}{c}{Methods} \\
\cmidrule(r){3-7}
 &  & \multicolumn{1}{c}{\begin{tabular}[c]{@{}c@{}}\small Megatron\end{tabular}} & \multicolumn{1}{c}{Merak} & \multicolumn{1}{c}{\begin{tabular}[c]{@{}c@{}} Cross-pass \\ schedule\end{tabular}} & \multicolumn{1}{c}{\begin{tabular}[c]{@{}c@{}}+ Fine-grained\\  recomputation\end{tabular}} & \multicolumn{1}{c}{\begin{tabular}[c]{@{}c@{}}+ Planner\\ (Oases)\end{tabular}} \\
\midrule
\multirow{3}{*}{\begin{tabular}[c]{@{}c@{}}NVLink\\ 3090\end{tabular}} & BERT-1.3B & 31.4 & 37.3 (1.19\(\times\)) & 41.2 (1.31\(\times\)) & 43.9 (1.40\(\times\)) & 50.9 (1.62\(\times\)) \\
 & LLaMA-7B  & 18.1 & 21.5 (1.18\(\times\)) & 22.6 (1.25\(\times\)) & 23.8 (1.31\(\times\)) & 25.2 (1.39\(\times\)) \\
 & LLaMA-65B  & 5.6 & 6.9 (1.25\(\times\)) & 7.7 (1.39\(\times\)) & 8.9 (1.60\(\times\)) & 9.1 (1.63\(\times\)) \\
\midrule
\multirow{3}{*}{3090} & BERT-1.3B & 23.0 & 29.2 (1.27\(\times\)) & 32.6 (1.42\(\times\)) & 42.5 (1.85\(\times\)) & 44.7 (1.95\(\times\)) \\
 & LLaMA-7B & 14.0 & 17.9 (1.28\(\times\)) & 22.4 (1.60\(\times\)) & 23.8 (1.69\(\times\)) & 24.3 (1.73\(\times\)) \\
 & LLaMA-65B & 5.3 & 6.2 (1.18\(\times\)) & 7.7 (1.46\(\times\)) & 8.9 (1.68\(\times\)) & 9.0 (1.71\(\times\)) \\
\bottomrule
\end{tabular}
\end{table*}

\subsection{Ablation study}\label{sec:evalablation}

We conduct an ablation study on diverse models and clusters to investigate the contributions of Oases’s optimizations, and the results are shown in Table~\ref{tab:ablation}.
Our cross-pass schedule provides stable accelerations, as the joint scheduling efficiently hides TMP communications.
The fine-grained recomputation provides additional acceleration in models with more communication constraints, such as the BERT-1.3B model on the 3090 cluster and the LLaMA-65B model on both clusters, where Megatron also exhibits lower device utilization.
Our planner performs better in models with more layers, as more model blocks bring a larger optimization space, bringing a more obvious acceleration. 
The parallel strategies are shown in the Table~\ref{tab:scheme}. 
The planner provides minor speedups on the LLaMA-65B model layers due to its larger data resharding overhead, which is covered marginally by performance gains from shrunk TMP degrees.

\begin{figure}
\centering
  \includegraphics[width=\linewidth]{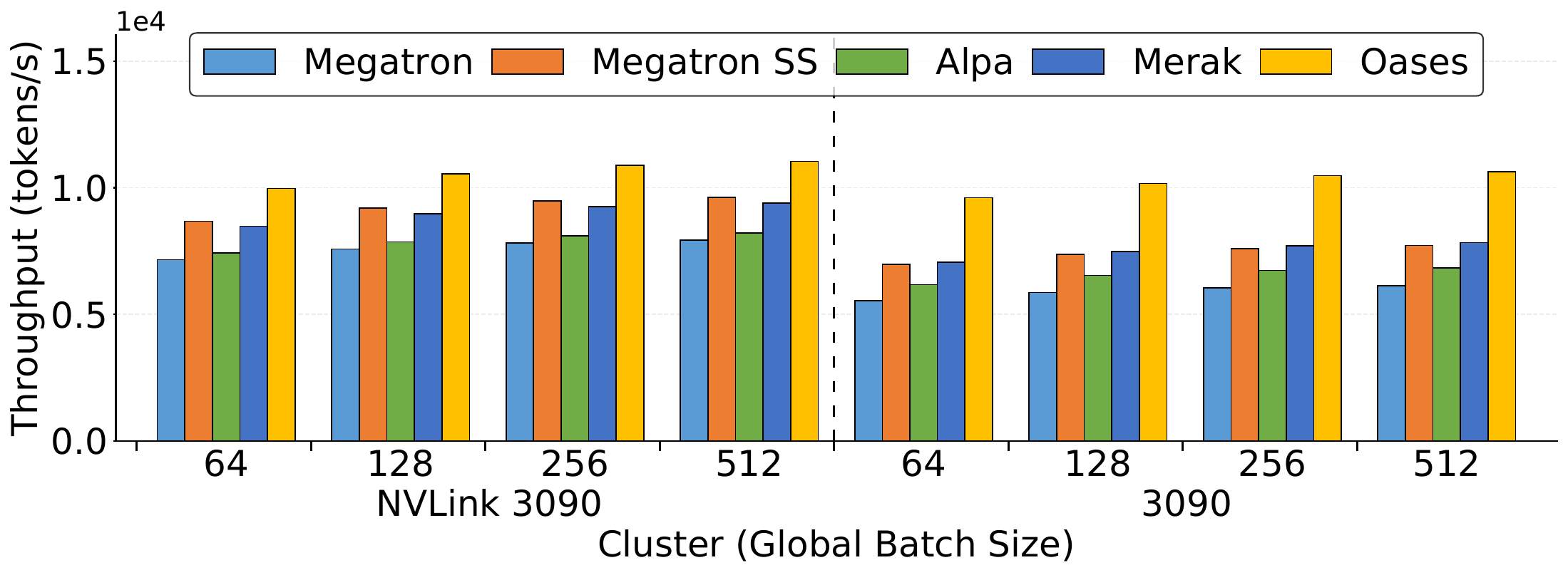}
\caption{Training performance on complete GPT-7.5B model with pipeline model parallelism. "Megatron SS" represents the Megatron with selective recomputation. Compared to Megatron SS, Oases can accelerate training by 1.15–1.38\(\times\).}
\label{fig:selectrecom}
\end{figure}

\subsection{Compare with other recomputation strategy}\label{apdx:costacc}
Oases requires the full recomputation strategy, recent optimizations come up with a selective recomputation strategy~\cite{korthikanti2023reducing}, which compares the memory-saving effectiveness of each component of transformer layers, and recomputes only the attention operation to reduce recomputation overhead. 
However, commodity servers are more memory-constrained, and trade memory for training performance is more costly.
Therefore, we add a comparison with Megatron that uses selective recomputation (denoted as "Megatron SS"). 
Note that Megatron SS will lead to out-of-memory errors when training GPT-18.4B model in Section~\ref{sec:evalpmp}, so we use GPT-7.5B model, following the model settings in Megatron~\cite{narayanan2021efficient}. 
The results are shown in Fig.~\ref{fig:selectrecom}. Megatron SS consumes the most GPU memory, becoming the fastest baseline on cluster NVLink 3090. Compared to Megatron SS, Oases can accelerate training by 1.15\(\times\) on cluster NVLink 3090 and 1.38\(\times\) on cluster 3090.

\subsection{Cost model accuracy}\label{apdx:costacc}

\begin{figure}
  \centering
  \begin{subfloat}[Cluster NVLink 3090.]{
  \includegraphics[width=0.48\linewidth]{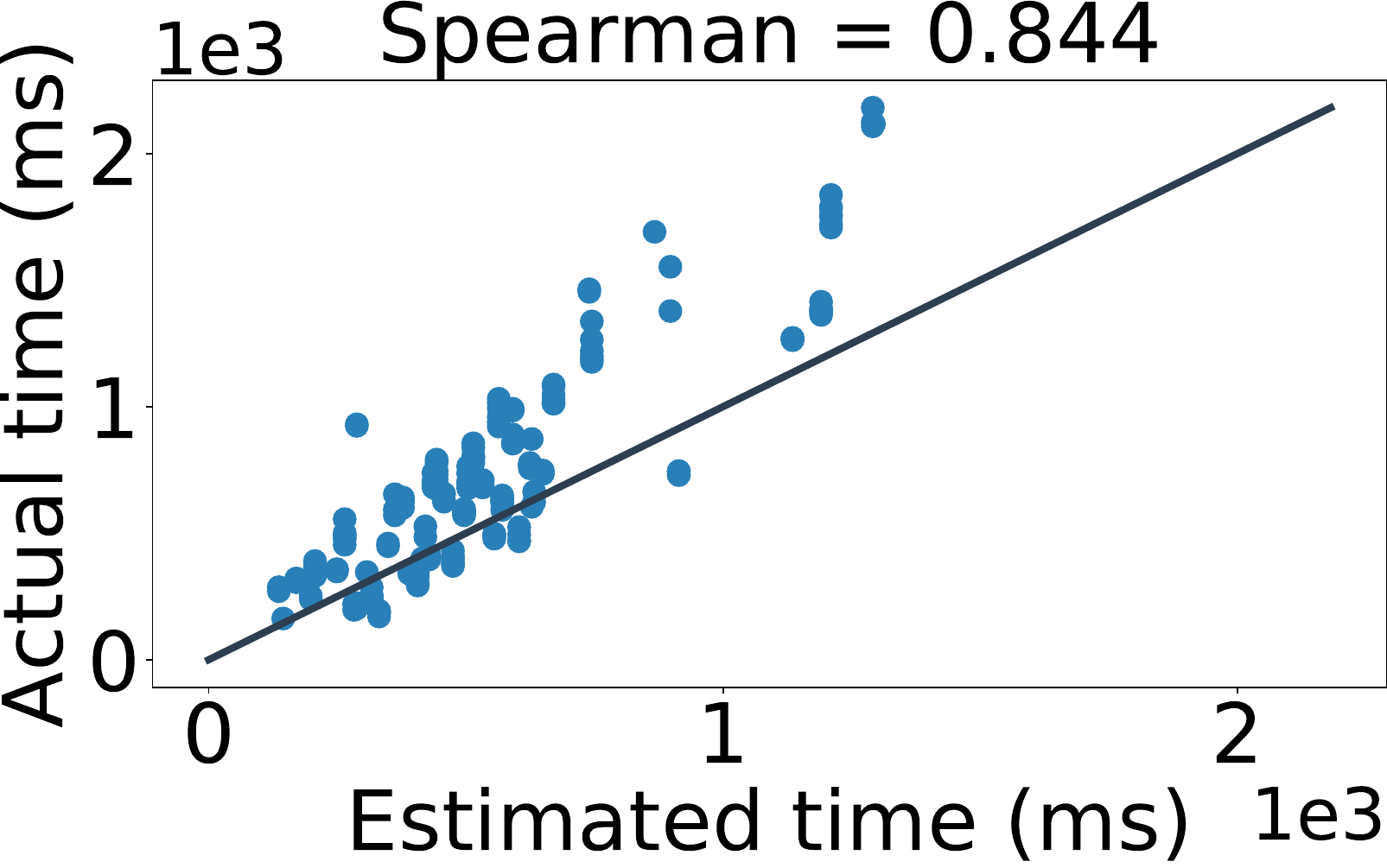}}
  \end{subfloat}
  \hfill
  \begin{subfloat}[Cluster 3090.]{
  \includegraphics[width=0.48\linewidth]{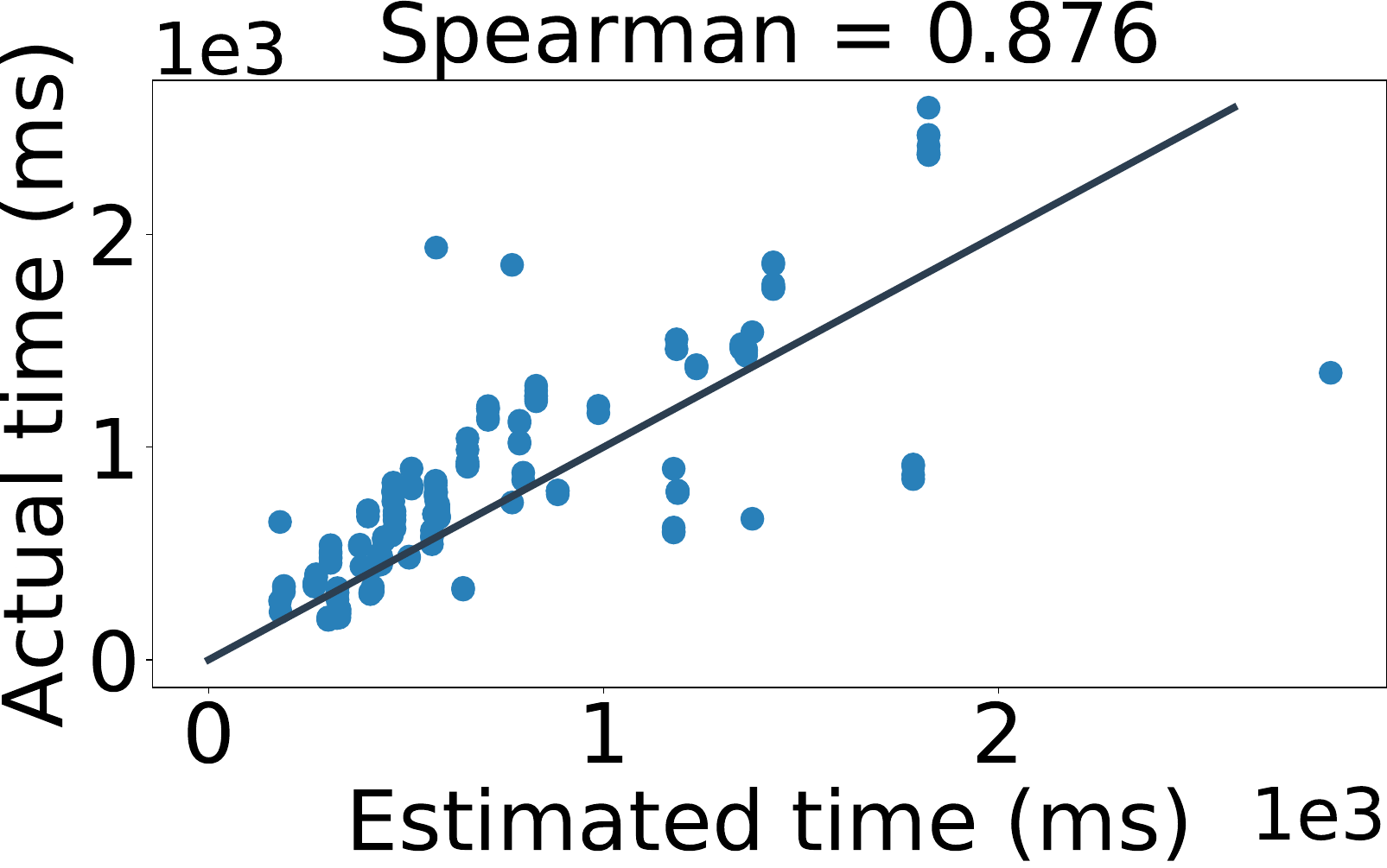}}
  \end{subfloat}
  \caption{Iteration time comparisons between the cost model estimate and actual cost. The Spearman correlation coefficient ranges from -1 to 1, with a value of 0.5 regarded as a relatively strong positive correlation.}
  \label{fig:costmodel}
\end{figure}

To evaluate the accuracy of Oases planner cost model in Section~\ref{sec:costmodel}, we compare the estimated iteration time from the cost model to the actual iteration time, and the results are shown in Fig.~\ref{fig:costmodel}.
Since the objective of Oases planner is to provide a strategy with the best performance,
we evaluate the precision of the cost model based on its efficiency in ranking different parallel strategies, with less emphasis on the absolute value of its prediction. 
The Spearman correlation results of 0.844 and 0.876 show a strong positive correlation between the outcome of cost model and the real situation.

\begin{figure}[tb]
  \centering
  \begin{subfloat}[BERT-large model.]{
  \includegraphics[width=0.475\linewidth]{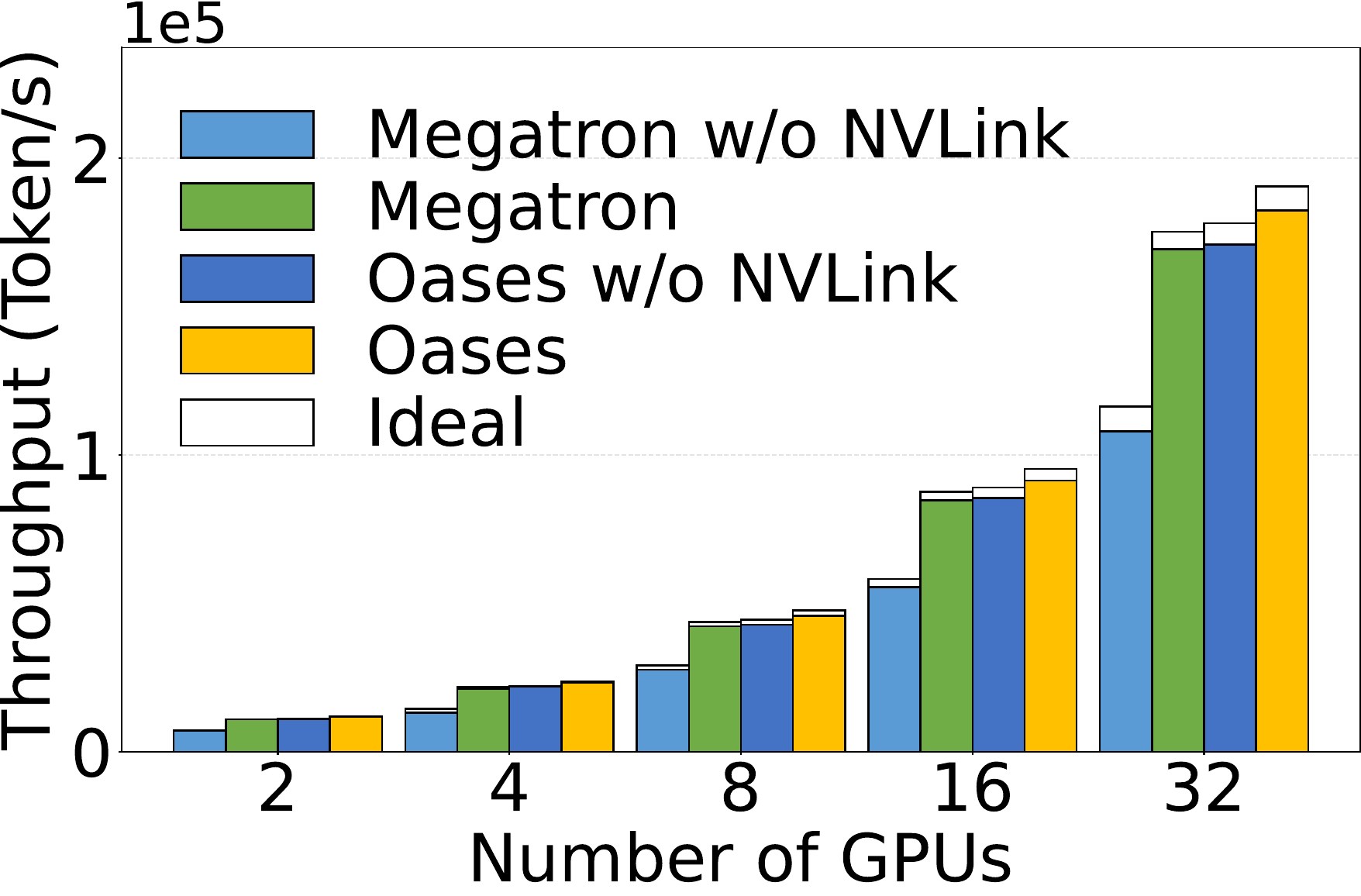}\label{fig:scalaa}}
  \end{subfloat}
  \hfill
  \begin{subfloat}[BERT-1.3B model.]{
  \includegraphics[width=0.475\linewidth]{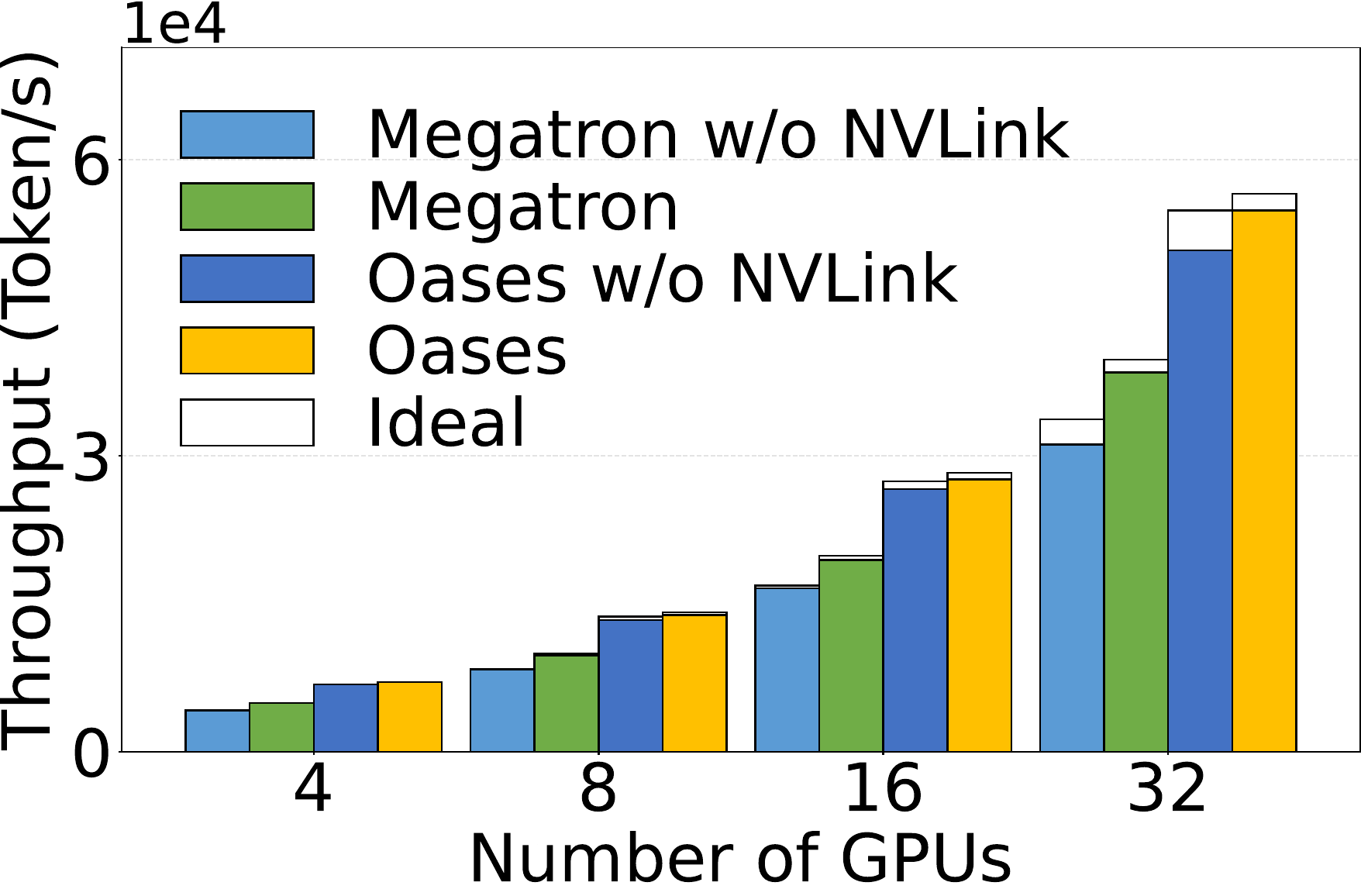}\label{fig:scalab}}
  \end{subfloat}
  
  \begin{subfloat}[GPT-8.3B model.]{
  \includegraphics[width=0.475\linewidth]{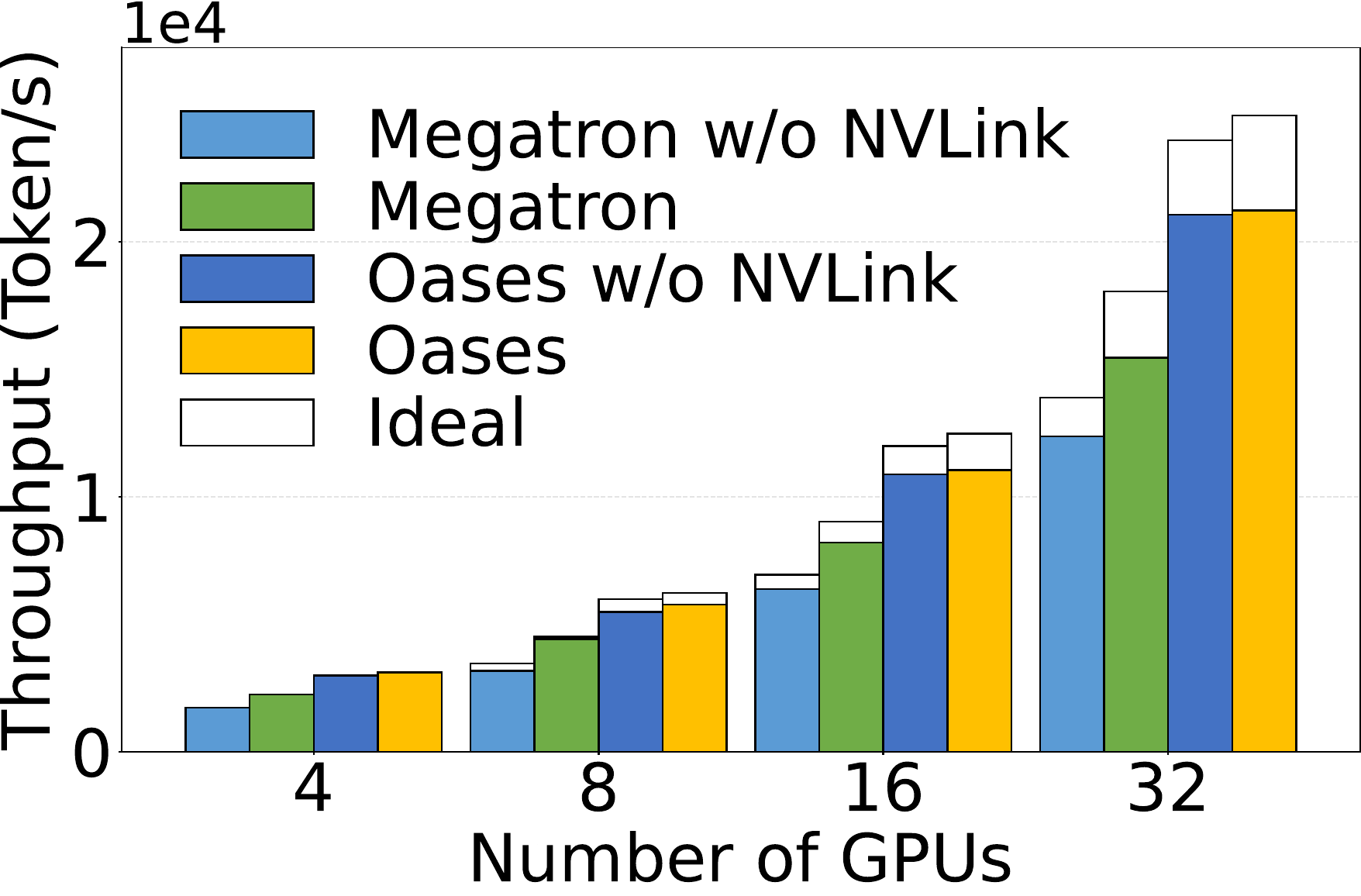}\label{fig:scalac}}
  \end{subfloat}
  \hfill
  \begin{subfloat}[LLaMA-7B model.]{
  \includegraphics[width=0.475\linewidth]{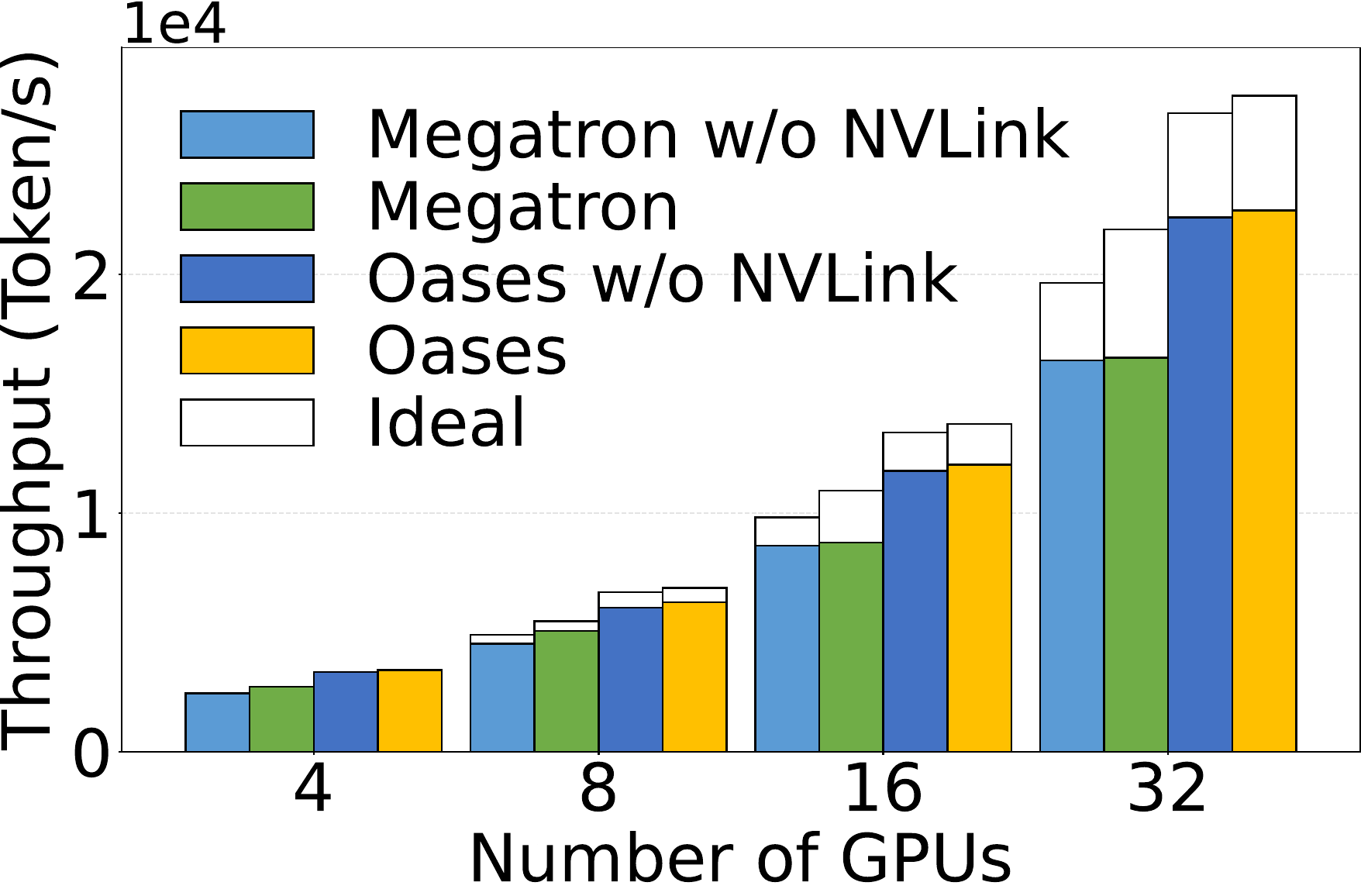}\label{fig:scalad}}
  \end{subfloat}
  \caption{Weak scaling performance compared to Megatron and related ideal linear results.}
  \label{fig:scala}
\end{figure}

\subsection{Scalability experiment}\label{apdx:scala}
We evaluate the scalability of Oases through weak scaling analysis, in which the global batch size and the number of GPUs are increased proportionally by adjusting the DP degree.
As shown in Fig.~\ref{fig:scala}, we compare the training throughputs between Oases and Megatron on both the 3090 and the NVLink 3090 clusters. We also compare training performance with corresponding ideal linear scaling baselines, computed by scaling the throughput from DP degree 1.
Both Megatron and Oases demonstrate good scalability, as DP efficiently scales training. 
The experimental results present trends of larger models and training with the NVLink 3090 cluster have worse scalability. 
This is primarily due to increased communication overhead associated with DP in these settings.
And the performance degradation between adjacent scales diminishes, which aligns with the theoretical communication scalability of AllReduce operations used in DP~\cite{patarasuk2009bandwidth}, given by \(2(N-1)/N\), where \(N\) is the DP degree. 
Due to the limited scope of our cluster, we cannot conduct tests on more GPUs. Nevertheless, given the good scalability of Oases, we expect it to continue delivering performance benefits as the number of devices increases.

\subsection{Performance on datacenter GPUs}\label{datacenter}
Oases is designed for TMP training on commodity servers with limited inter-device connections. 
However, some older datacenter GPUs are equipped with high-speed inter-device connections and can offer competitive cost-effectiveness compared to commodity GPUs.
To evaluate Oases in this scenario, we conduct additional experiments with datacenter GPUs NVIDIA V100 connected by NVLinks. 
Each V100 has 32GB memory compared to the 24GB 3090 GPU, so we increase the number of transformer layers accordingly. The experimental results are shown in Table VI and demonstrate that Oases achieves 1.07--1.10\(\times\) end-to-end training speedup compared to Megatron. 
Although the overall speedup is modest due to the low communication-to-computation ratio in the fully connected NVLink environment, Oases substantially reduces communication volume by 5.46--7.97\(\times\). 
We anticipate that Oases will yield even greater benefits on more advanced GPUs with higher computational performance, where the communication cost during TMP training becomes more significant.

\begin{table} \renewcommand\arraystretch{1} 
\small
\setlength\tabcolsep{2pt}
\centering
\caption{End-to-end performance on 8 V100 GPUs which are fully connected via NVLinks. The communication ratio refers to the ratio of non-overlapping communication.}
\label{tab:v100}
\resizebox{\linewidth}{!}{
\begin{tabular}{@{}ccccccc@{}}
\toprule
\multirow{2}{*}{\begin{tabular}[c]{@{}c@{}}Model layer \\ name\end{tabular}} &
  \multirow{2}{*}{\begin{tabular}[c]{@{}c@{}}\# of \\ layers\end{tabular}} &
  \multirow{2}{*}{\begin{tabular}[c]{@{}c@{}}TMP \\ degree\end{tabular}} &
  \multicolumn{2}{c}{Throughput(k tokens/s)} &
  \multicolumn{2}{c}{Comm. ratio} \\ \cmidrule(l){4-7} 
      &    &   & Megatron & Oases          & Megatron & Oases \\ \midrule
BERT-large& 30 & 2 & 224.5    & 245.8 (1.10\(\times\)) & 16.33\%   & 2.99\%  \\
BERT-1.3B& 30 & 4 & 90.0     & 99.4 (1.10\(\times\))  & 14.94\%    & 2.36\%  \\
GPT-8.3B& 30 & 4 & 50.6     & 55.7 (1.10\(\times\))  & 14.20\%    & 1.78\%  \\
LLaMA-7B& 20 & 4 & 47.4     & 51.9 (1.10\(\times\))  & 13.43\%    & 2.31\%  \\
GPTNeoX& 20 & 8 & 25.0     & 27.3 (1.09\(\times\))  & 12.34\%    & 1.73\%  \\
LLaMA-65B& 10 & 8 & 31.5     & 34.3 (1.09\(\times\))  & 11.68\%    & 1.14\%  \\
GPT-3& 6  & 8 & 20.8     & 22.2 (1.07\(\times\))  & 8.23\%     & 1.21\%  \\ \bottomrule
\end{tabular}}
\end{table}

\section{Related works}\label{apdx:relatedwork}
\noindent\textbf{Large-scale model training.}
Researchers use a variety of parallel methods for large-scale model training. Data parallelism~\cite{li2020pytorch,sergeev2018horovod} is commonly used but restricted by the device capacity.
Recent works rematerialize runtime memories~\cite{chen2016training,jain2020checkmate,sun2022stronghold} and reduce the redundant memory across data-parallel devices~\cite{rajbhandari2020zero,zhao2023pytorch,zhou2023mpress}. 
While model parallelism scatters models into device groups. Pipeline model parallelism~\cite{huang2019gpipe,liu2022autopipe} divides model layers into stages and executes them in a pipeline.
Tensor model parallelism (TMP)~\cite{shoeybi2019megatron,wang2022tesseract,karakus2021amazon} partitions model operators and drastically reduces the memory usage but results in high communication overhead. A combination of these methods is often necessary for large-scale model training~\cite{jia2022whale,zheng2022alpa,unger2022unity,lai2023merak,li2023colossal} and can complement our work since Oases focuses on the communication-computation overlap of TMP.

\noindent\textbf{Automatic model partitioning.}
Recent studies have proposed automated methods for optimizing model partitioning. For pipeline model parallelism, works use algorithms including heuristic algorithm~\cite{liu2022autopipe} and dynamic programming~\cite{fan2021dapple,narayanan2019pipedream,li2021terapipe} to balance the overhead between pipeline stages. For model-parallel training, researchers search for partition schemes with methods such as randomized MCMC~\cite{jia2019beyond}.
Finding an optimal parallel strategy for multiple parallel methods together will greatly enlarge the optimization space. Prior works~\cite{li2022amp, tarnawski2021piper, jia2022whale, unger2022unity, zheng2022alpa, miao2022galvatron} mainly focus on how to simplify and solve this problem.
The training schedule of Oases can be applied to the model-parallel methods to achieve efficient training. And existing automatic model partition methods cannot apply to overlapped TMP training, which is the core problem Oases planner tries to answer. 

\noindent\textbf{Communication-computation overlap.}
Overlapping is an effective technique to reduce communication overhead. 
To reduce the gradient aggregation overhead in data-parallel training, communication scheduling~\cite{zhang2017poseidon,shi2019mg,peng2019generic,li2022embrace,mahajanSYNDICATE} is intensively studied. 
In pipeline model parallelism, peer-to-peer communications happen between pipeline stages, and methods~\cite{oh2022out,ye2021hippie,zhuang2023optimizing} focus on overlapping this communication with computation.
Works on data parallelism and pipeline model parallelism are orthogonal to Oases since TMP is generally combined with them.
The communication of TMP holds a larger proportion in large-scale model training, approaches~\cite{wang2022overlap,zeng2022acctfm,lai2023merak,jangda2022breaking} overlap the communications with decoupled operations or data.
And Oases propose a more exhaustive overlapped training schedule for TMP in this paper.

\section{Conclusion}
We present Oases in this paper, a training approach that can overlap communication and computation of TMP training and find the best model parallel strategy upon the overlapped training schedule. 
Oases can accelerate model-parallel training by 1.01--1.63\(\times\) in a NVLink cluster, and by 1.20--1.95\(\times\) in 3090 servers. Additionally, Oases sees an acceleration in complete large-scale model training of up to 1.72\(\times\) when combined with PMP.
We currently implement and evaluate Oases only on transformer-based models. The ideas behind this work could be easily extended to other model-parallel powered training, which will be our future work.
We hope Oases will be used with other distributed methods to democratize large-scale model training in commodity hardware.

\bibliographystyle{IEEEtran}
\bibliography{IEEEabrv,sample}


 




\vspace{-10 mm}
\begin{IEEEbiography}[{\includegraphics[width=1in,height=1.25in,clip,keepaspectratio]{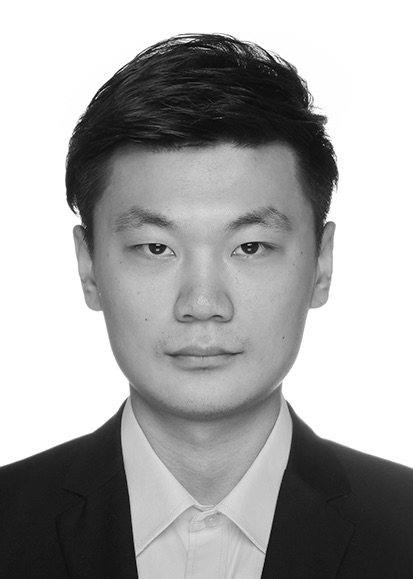}}]{Shengwei Li}
received the B.S. degree from Nanjing University, Jiangsu, China in 2017, and the M.S. degree in computer science from Stony Brook University, New York, USA in 2020. He is pursuing his Ph.D. degree at the College of Computer Science and Technology, NUDT. His research interests include high-performance computing and distributed machine learning systems.
\end{IEEEbiography}

\vspace{-10 mm}
\begin{IEEEbiography}[{\includegraphics[width=1in,height=1.25in,clip,keepaspectratio]{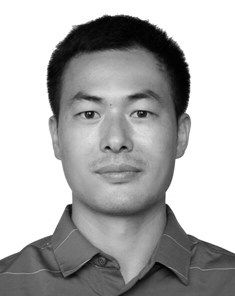}}]{Zhiquan Lai}
received his Ph.D, M.S. and B.S. degrees in Computer Science from National University of Defense Technology (NUDT) in 2015, 2010 and 2008 respectively.
He is currently an associate professor in the National Key Laboratory of Parallel and Distributed Computing of NUDT.
He worked as a research assistant at Department of Computer Science, the University of Hong Kong during Oct. 2012 to Oct. 2013. His current research interests include high-performance system software, distributed machine learning, and power-aware computing.
\end{IEEEbiography}

\vspace{-10 mm}
\begin{IEEEbiography}[{\includegraphics[width=1in,height=1.25in,clip,keepaspectratio]{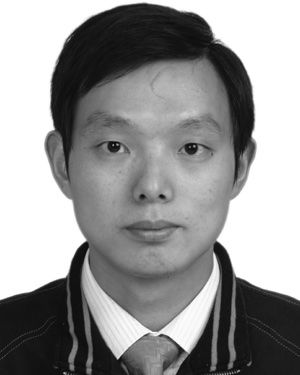}}]{Dongsheng Li}
is a professor and doctoral supervisor in the College of Computer Science and Technology at National University of Defense Technology (NUDT). He received his PhD degree in computer science and technology from NUDT in 2005. He was awarded the Chinese National Excellent Doctoral Dissertation in 2008. His research interests include distributed systems, cloud computing, and big data processing.
\end{IEEEbiography}

\vspace{-10 mm}
\begin{IEEEbiography}[{\includegraphics[width=1in,height=1.25in,clip,keepaspectratio]{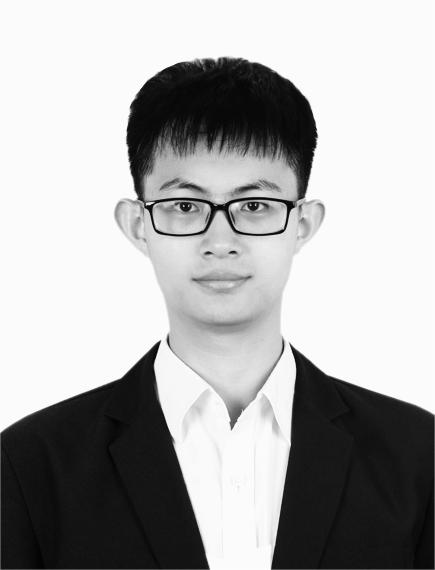}}]{Yanqi Hao} 
received his Bachelor degree in software engineering from Northwestern Polytechnical University, China, in 2022. He is pursuing his M.S. degree at the College of Computer Science and Technology, NUDT. His current interests are mainly in optimization techniques related to large-scale model training.
\end{IEEEbiography}

\vspace{-10 mm}
\begin{IEEEbiography}[{\includegraphics[width=1in,height=1.25in,clip,keepaspectratio]{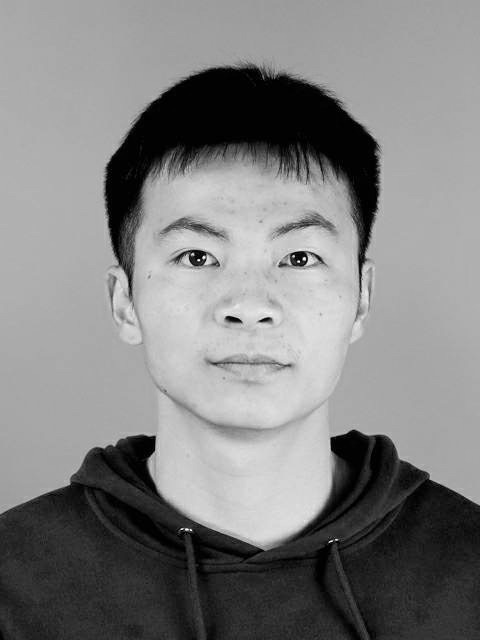}}]{Weijie Liu}
received his Bachelor degree in computer science from Nankai University, China, in 2020, and his M.S. degree from the College of Computer Science and Technology, National University of Defense Technology (NUDT), in 2022. He is pursuing his Ph.D. degree at the College of Computer Science and Technology, NUDT. His current interests are mainly in optimization techniques related to large-scale model training.
\end{IEEEbiography}

\vspace{-10 mm}
\begin{IEEEbiography}[{\includegraphics[width=1in,height=1.25in,clip,keepaspectratio]{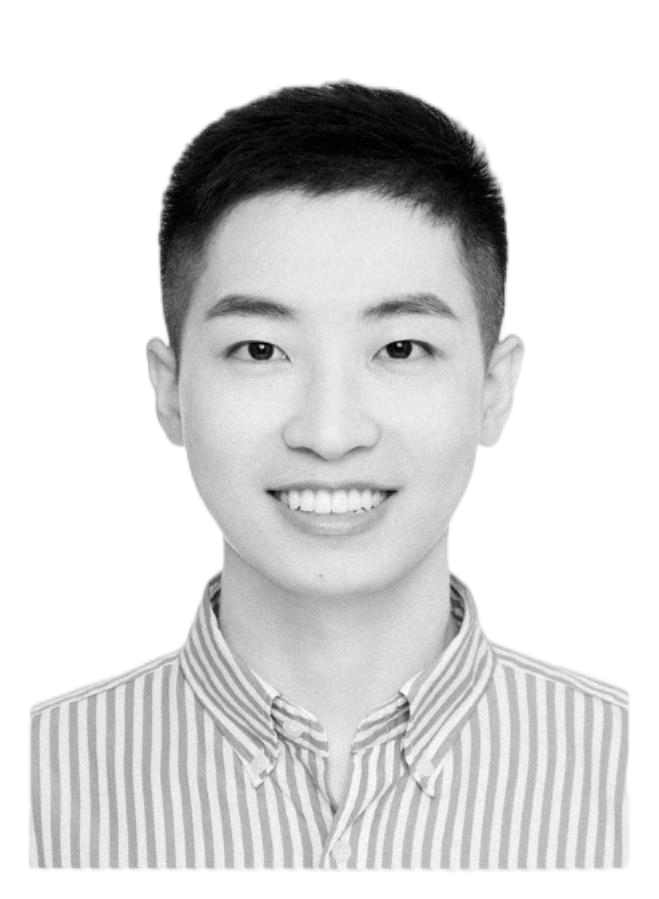}}]{Keshi Ge} received his B.S. degree from the Department of Computer Science and Technology, Xi’an Jiaotong University, China, in 2015, and his Ph.D. and M.S. degree from the College of Computer Science and Technology, National University of Defense Technology (NUDT), in 2022 and 2017, respectively. He worked as a visiting Ph.D. student at the Department of Electrical and Computer Engineering, University of Alberta, from Nov. 2019 to Aug. 2020. He is currently an Assistant Professor with NUDT. His research interests include high-performance computing and distributed machine learning systems.
\end{IEEEbiography}

\vspace{-10 mm}
\begin{IEEEbiography}[{\includegraphics[width=1in,height=1.25in,clip,keepaspectratio]{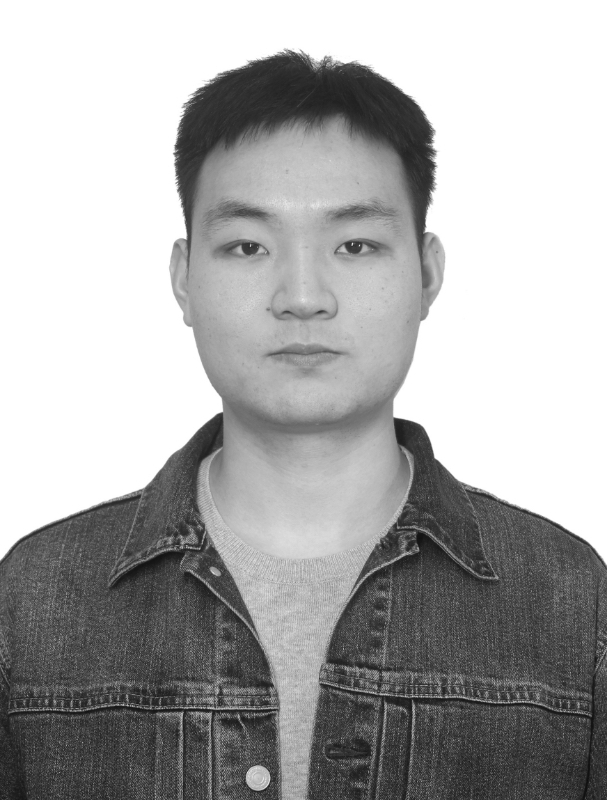}}]{Xiaoge Deng}received the B.S. degree in Mathematics from the University of Science and Technology of China (USTC), Hefei, China, in 2018. He obtained the M.S. and Ph.D. degrees in Computer Science from the College of Computer Science and Technology at the National University of Defense Technology (NUDT), Changsha, China, in 2020 and 2024, respectively. Currently, he is an Assistant Professor at the Intelligent Game and Decision Lab in Beijing, China. His research interests include optimization algorithms and distributed machine learning. 
\end{IEEEbiography}

\vspace{-10 mm}
\begin{IEEEbiography} [{\includegraphics[width=1in,height=1.25in,clip,keepaspectratio]{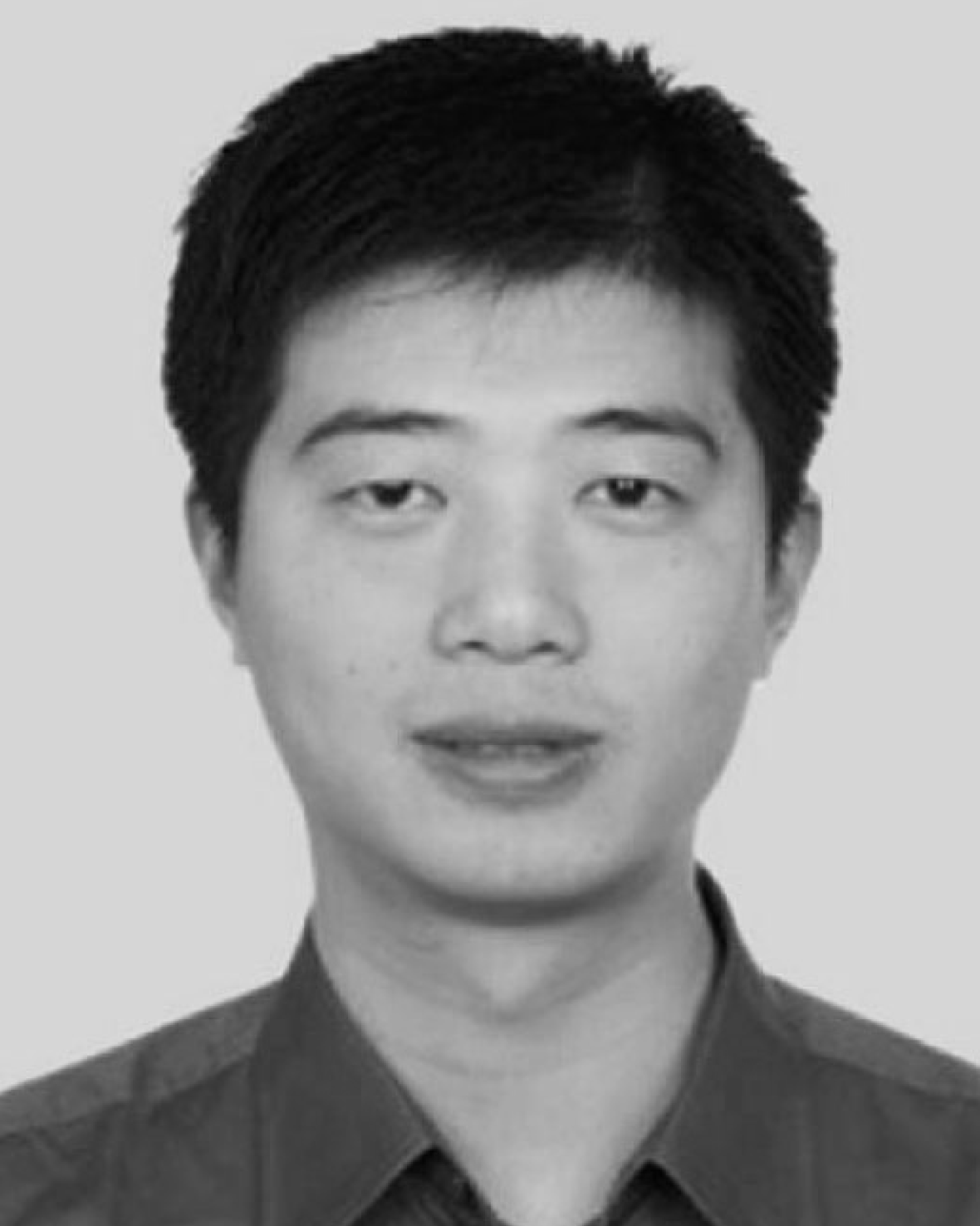}}] 
{Kai Lu} received the BS and Ph.D. degrees from the National University of Defense Technology in 1995 and 1999, respectively. He is now a College of Computer Science and Technology, National University of Defense Technology professor. His research interests include parallel programming and operating system and security.
\end{IEEEbiography}

\vfill

\end{document}